\documentclass[reprint,notitlepage,twocolumn,
superscriptaddress,aps,longbibliography]{revtex4-1}
\usepackage[utf8]{inputenc}
\usepackage{braket}
\usepackage{amsmath}
\usepackage{amsthm}
\usepackage{mathrsfs}
\usepackage{mathtools}
\usepackage{amssymb}
\usepackage{dsfont}
\usepackage{braket}
\usepackage{bm}
\DeclareMathOperator{\tr}{tr}
\usepackage{color}
\pdfoutput=1

\usepackage[english]{babel}
\usepackage[T1]{fontenc}
\usepackage{amsmath}
\usepackage[colorlinks=true]{hyperref}
\usepackage{enumitem}
\usepackage{amsthm}
\usepackage{amssymb}
\usepackage{graphicx}
\usepackage{float}
\usepackage[ruled,vlined]{algorithm2e}

\begin{document}
\title{ 
Flexible learning of quantum states with generative query neural networks }

\author{Yan Zhu$^1$, Ya-Dong Wu$^{1,*}$, Ge Bai$^1$, Dong-Sheng Wang$^2$, Yuexuan Wang$^{1,3}$ and Giulio Chiribella$^{1,4,5,\dagger}$\\
$^1$QICI Quantum Information and Computation Initiative, Department of Computer Science,
The University of Hong Kong, Pokfulam Road, Hong Kong.\\
$^2$CAS Key Laboratory of Theoretical Physics, Institute of Theoretical Physics, Chinese Academy of Sciences, Beijing 100190, People’s
Republic of China. \\
$^3$College of Computer Science and Technology, Zhejiang University, Hangzhou, China. \\
$^4$Department of Computer Science, Parks Road, Oxford, OX1 3QD, United Kingdom. \\
$^5$Perimeter Institute for Theoretical Physics, Waterloo, Ontario N2L 2Y5, Canada. \\
These authors contributed equally: Yan Zhu, Ya-Dong Wu. \\
Emails: $^*$yadongwu@hku.hk,
$^\dagger$giulio@cs.hku.hk}

\iffalse
\author{Yan Zhu}
\thanks{These authors contributed equally}
\affiliation{QICI Quantum Information and Computation Initiative, Department of Computer Science,
The University of Hong Kong, Pokfulam Road, Hong Kong}
\author{Ya-Dong Wu}
\email{yadongwu@hku.hk}
\thanks{These authors contributed equally}
\affiliation{QICI Quantum Information and Computation Initiative, Department of Computer Science,
The University of Hong Kong, Pokfulam Road, Hong Kong}
\author{Ge Bai}
\affiliation{QICI Quantum Information and Computation Initiative, Department of Computer Science,
The University of Hong Kong, Pokfulam Road, Hong Kong}
 \author{Dong-Sheng Wang}
\affiliation{CAS Key Laboratory of Theoretical Physics, Institute of Theoretical Physics, Chinese Academy of Sciences, Beijing 100190, People’s
Republic of China}
\author{Yuexuan Wang}
\affiliation{QICI Quantum Information and Computation Initiative, Department of Computer Science,
The University of Hong Kong, Pokfulam Road, Hong Kong}
\affiliation{College of Computer Science and Technology, Zhejiang University, Hangzhou, China}
\author{Giulio Chiribella}
\email{giulio@cs.hku.hk}
\affiliation{QICI Quantum Information and Computation Initiative, Department of Computer Science,
The University of Hong Kong, Pokfulam Road, Hong Kong}
\affiliation{Department of Computer Science, Parks Road, Oxford, OX1 3QD, United Kingdom}
\affiliation{Perimeter Institute for Theoretical Physics, Waterloo, Ontario N2L 2Y5, Canada}
\fi

\begin{abstract}
  Deep neural networks are a powerful tool for the characterization of quantum states. 
   Existing  networks  are typically trained with experimental data gathered from the specific quantum state that needs  to be characterized. 
 But is it possible to train a neural network  offline and to make   predictions about  quantum states other than the ones used for  the  training? 
 Here we introduce a model of network  that  can be trained with classically simulated data from a fiducial set of states and measurements, and can  later be used to  characterize  quantum states that share structural similarities with the states in  the fiducial set. With little guidance of quantum physics,  the network  builds 
 its own data-driven   representation of quantum states, and then uses it to predict the outcome statistics of quantum measurements  that have not been performed yet.
 The state representation produced by the network can also  be used for tasks beyond the prediction of outcome statistics, including  clustering of quantum states and  identification of different phases of matter. 
 Our network model provides a flexible  approach that can be applied to  online learning  scenarios, where predictions must be generated  as soon as experimental data become available,  and to  blind learning scenarios where  the learner has only access to an encrypted description of the quantum hardware.
  \end{abstract}

\maketitle

\section{Introduction}\label{sec:intro}
Accurate characterization of quantum  hardware  is crucial for the development, certification, and benchmarking of new quantum technologies~\cite{eisert2020}. Accordingly, major efforts have been invested into developing suitable techniques for characterizing quantum states, including  quantum state tomography~\cite{toth2010,gross2010,cramer2010,lanyon2017,Cotler2020}, classical shadow estimation~\cite{huang2020,huang2021}, partial state characterization~\cite{PhysRevLett.106.230501,PhysRevLett.107.210404} and quantum state learning~\cite{aaronson2007,aaronson2019,aaronson2019online,arunachalam2020}.
Recently, the dramatic development of artificial intelligence inspired new approaches on machine learning methods~\cite{carleo2019}.  
 In particular, a sequence of works explored applications of neural networks to various state characterization tasks~\cite{torlai2018np,Torlai2018prl,xu2018,carrasquilla2019,tiunov2020,Ahmed2021PRL,Ahmed2021PRR,rocchetto2018,quek2021,palmieri2020,smith2021}.
%especially, using generative models to learn quantum states from samples of measurement outcomes~\cite{torlai2018np,Torlai2018prl,rocchetto2018,carrasquilla2019}. 

In the existing quantum applications,  neural networks are typically trained using experimental data generated from  the specific  quantum state that needs to be characterized. 
As a consequence, the information learnt in the training phase cannot be directly transferred to other states: for a new quantum state, a new training procedure must be carried out. 
This structural limitation affects the learning efficiency  in applications involving  multiple quantum states, including important tasks such as quantum state clustering~\cite{gael2019}, quantum state classification~\cite{gao2018}, and quantum cross-platform verification~\cite{elben2020}. 
%which require comparisons of  quantum states.
%restricts the applicability to scenarios involving a single unknown quantum state, and %Furthermore, a neural network trained for a particular chosen set of POVM measurements cannot be applied to a measurement dataset coming from a different set of POVM measurements. {\color{blue} cannot understand...}

 In this paper, we develop a flexible model of neural network  that can be trained offline using simulated data from a fiducial set of states and measurements, and is capable of learning  multiple quantum states that share structural similarities   with the fiducial states, such as being ground states in the same  phase of a quantum manybody system.     %, after training, can reproduce multiple measurement outcomes {\color{blue} [I am not sure that "reproduce multiple measurement outcomes" is the appropriate expression here]} of various quantum states on demand.
Our model, called generative query network for quantum state learning (GQNQ), takes advantage of a technique originally  developed in classical image processing for learning 3D scenes from 2D snapshots taken from different viewpoints~\cite{eslami2018}.  
%GQNQ  combines two major  techniques in deep learning: representation learning~\cite{bengio2013} and the generative model~\cite{foster2019}. 
 The key idea is to use a representation network~\cite{bengio2013} to construct a data-driven representation of quantum states, and then  to feed this  representation into a generation network~\cite{foster2019} that predicts the outcome statistics of quantum measurements that have not been performed yet. The state representations produced by GQNQ  enable applications where multiple states have to be compared, such as quantum state clustering or the identification of different phases of matter. The applications of GQNQ are illustrated with numerical experiments on multiqubit states, including ground states of Ising models and XXZ models, and   continuous-variable quantum states, both Gaussian and non-Gaussian. 
   
The deep learning techniques developed in this work can be applied to real-time control and calibration of various quantum state preparation devices.   They can also be applied to online learning scenarios wherein predictions have to be  made as soon as data become available,  and to blind learning scenarios where the learner has to predict the behaviour of a quantum hardware without having access to its quantum  description, but only to an encrypted parametrization.

%The latter scenario is suitable for  applications where one party holds the quantum hardware and another party holds classical computational resources for data analysis.  Since GQNQ  does not require the description of the devices to be provided in clear, it can be used to perform the data analysis on a public server, without revealing the mapping between the encrypted parameters and the actual devices.  

\section{Results}

{\bf Quantum state learning framework.} \label{subsec:overview}   In this work we adopt a learning framework  inspired by the task of  ``pretty good tomography''~\cite{aaronson2007}.  An experimenter has a source that produces quantum systems in some unknown quantum state $\rho$. The experimenter's goal is to characterize $\rho$, becoming able  to make  predictions on the outcome statistics of a set of measurements of interest, denoted by $\mathcal M$. Each measurement $\bm M  \in  \cal M$ corresponds to a  positive operator-valued measure (POVM), that is, a set of positive operators  ${\bm M}:=\left  ( M_j\right)_{j=1}^k$ acting on the system's Hilbert space  and satisfying the normalization condition $\sum_{j=1}^k M_j =\mathds{1}$ (without loss of generality, we assume that all the measurements in $\cal M$ have the same number of  outcomes, denoted by $k$).  

To characterize the state $\rho$, the experimenter performs a finite number of measurements  $\bm{M}_i$, $i\in\{1,\dots,  s\}$,   picked at random from $\cal M$.  This random subset of measurements  will be denoted by ${\cal S} = \{  \bm{M}_i\}_{i=1}^s$.      Note that in general both $\mathcal M$ and $\cal S$ may not be informationally complete.

Each measurement in  $\mathcal S$ is performed multiple times on independent copies of the quantum state $\rho$, obtaining a vector of experimental frequencies  $\bm{p}_i$.        Using this data, the experimenter attempts   to  predict the outcome statistics of a new, randomly chosen measurement $\bm{M}' \in \mathcal M \setminus \mathcal S$.
For this purpose, the experimenter uses the assistance of an automated learning system (e.g. a neural network), hereafter called the learner. For each measurement  $\bm{M}_i\in \mathcal S$, the experimenter provides the learner with a pair $(\bm{m}_i, \bm{p}_i)$, where  $\bm{m}_i$  is a parametrization of the measurement  $\bm{M}_i$, and  $\bm{p}_i
$ is the vector of  experimental frequencies for the measurement $\bm{M}_i$.
%that has entries equal to the  frequencies  of the measurement outcomes. 
%In the limit of infinite statistics, the entries of the vector converge to the values $v_i^{(j)}  = \tr\left(\rho M_i^{(j)}\right)$ given by the Born rule for the state $\rho$ under consideration.  
Here the parametrization $\bm{m}_i$ could be the full description of the POVM  $\bm{M}_i$, or a lower-dimensional parametrization valid only for measurements in the set $\mathcal M$.   For example,  if $\cal M$ contains  measurements of linear polarization, a measurement in $\cal M$ could be parametrized by the angle $\theta$ of the corresponding polarizer. The parametrization could also be encrypted, so that the actual description of the quantum hardware in the experimenter's laboratory is  concealed from the learner. 

To obtain a prediction for a new, randomly chosen measurement  $\bm{M}'\in \mathcal M \setminus \mathcal S$, the experimenter sends its parametrization $\bm{m}'$ to the learner.  The learner's task  is to predict the correct outcome probabilities  $ \bm{p}'_{\text{true}}   =    \left(  \tr\left(\rho M'_j  \right)\right)_{j=1}^k$.    This task includes as special case quantum state reconstruction, corresponding to the situation where the subset $\mathcal S$ is informationally complete.

Note that, {\em a priori}, the learner may have no knowledge about quantum physics whatsoever. The ability to make reliable predictions about the statistics of quantum measurements  can be  gained  automatically through a training phase, where the learner is presented with  data  and adjusts its internal parameters  in a data-driven way.  In previous works~\cite{torlai2018np,Torlai2018prl,carrasquilla2019,tiunov2020,quek2021,smith2021}, the training was based on experimental data gathered from the same state $\rho$ that needs to be characterized. In the following, we will provide a  model of learner that can be trained with data from a fiducial set of quantum states that share some common structure with $\rho$,  but can generally be different from $\rho$. 
%such as belonging to the same quantum phase of a manybody quantum system. 
%The data  can be generated either from real experiments or from computer simulations.   
The density matrices of the fiducial states  can be completely unknown to the learner.  In fact, the learner does not even need to be provided a parametrization of the fiducial states:  the only piece of information that the learner needs to know is  which measurement data correspond to the same state.

\medskip

 {\bf The GQNQ network.}   Our model of learner, GQNQ, is a neural network composed of two main parts:  a representation network~\cite{bengio2013}, producing  a data-driven representation of quantum states, and a generation network~\cite{foster2019}, making predictions about the outcome probabilities of quantum measurements that have not been performed yet.   The combination of a representation network and a generation network is called a  generative query network~\cite{eslami2018}.  This type of neural  network was originally  developed for the classical task of  learning 3D scenes from 2D snapshots taken from different viewpoints. The intuition for adapting  this  model to the quantum domain is that the statistics of a fixed quantum measurement can be regarded as a lower-dimensional projection of a higher-dimensional object (the quantum state), in a way that is analogous to  a  2D projection of a 3D scene. The numerical experiments reported in this paper indicate that this intuition is indeed correct, and that  GQNQ works well even in the presence of errors in the measurement data and fluctuations due to finite statistics.  
 %In the original image processing application, the strength of generative query networks was their ability to generate 2D images from viewpoints that were not included in the original dataset. By analogy,  a generative query network for quantum state learning may be expected to be  good at  predicting the statistics of quantum measurements  that were not performed in the laboratory.    

%\begin{widetext}

\begin{figure*}
    \centering
    \includegraphics[width=0.7\textwidth]{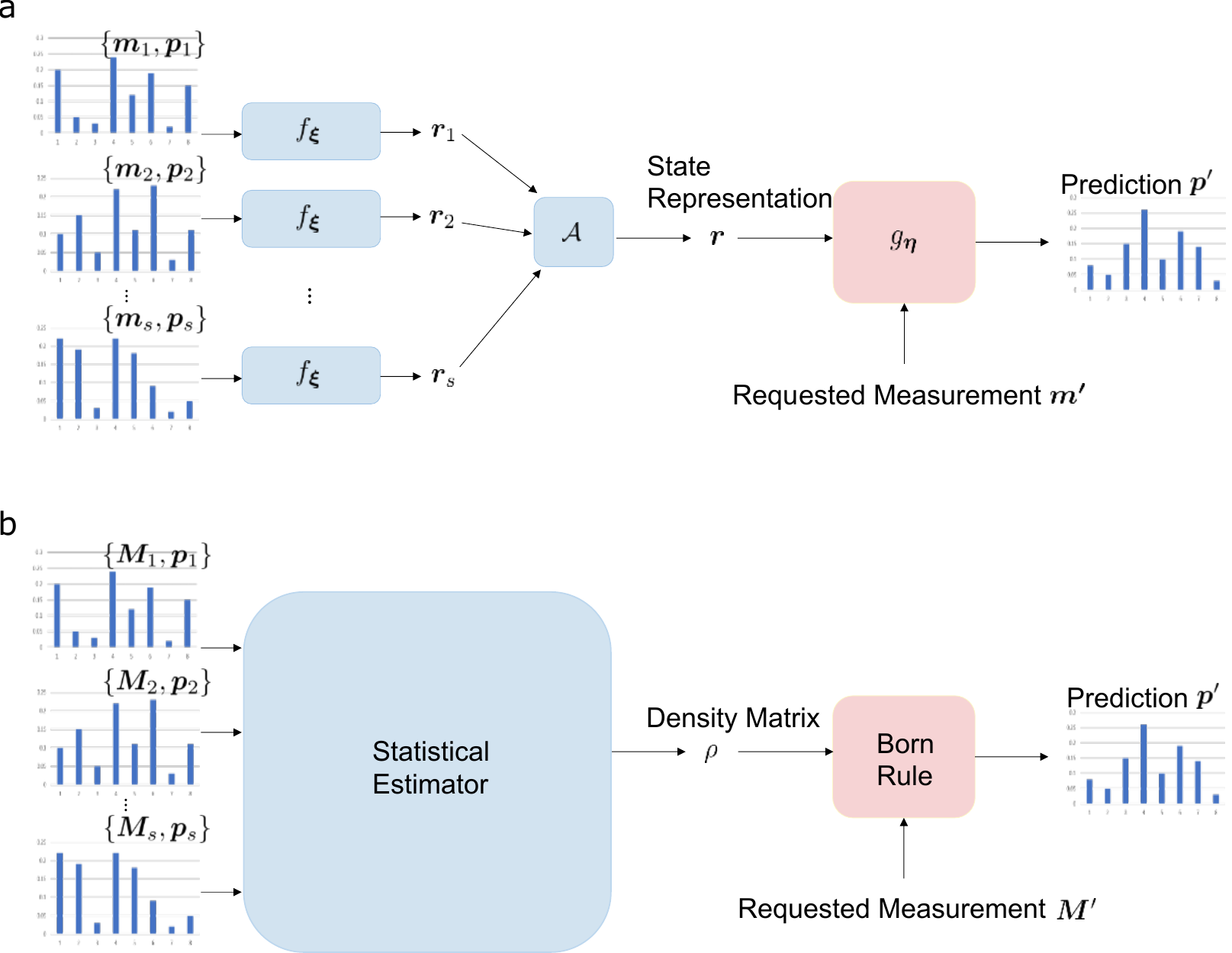}
    \caption{%{\color{red} add a subfigure describing RBM tomography}
   Structure of  GQNQ and comparison with  quantum state tomography.   
    In GQNQ (a), a representation network   receives as input the raw measurement data $ \{ (  \bm{m_i},  \bm{p}_i)\}_{i=1}^s$ and produces as output $s$ vectors $\bm{r}_i  =  f_{\bm \xi}  (  \bm{m_i},  \bm{p}_i)$, that are combined into a single vector  $\bm r$ by an aggregate function $\cal A$. The vector $\bm{r}$ serves as a concise representation of the quantum state, and is sent to a generation network $g_{\bm \eta}$, which predicts the outcome statistics  $\bm{p}'$ of any desired measurement $\bm{m}'$ in the set of measurements of interest.       In quantum  tomography (b),  the raw measurement data are fed into a statistical estimator (such as maximum likelihood), which produces a guess for the density matrix $\rho$. Then, the density matrix is used to predict  the outcome probabilities of unperformed quantum measurements via the Born rule.  Both GQNQ and quantum tomography use  data to infer  a representation of the quantum state.  
    %The main differences are that GQNQ (1) does not assume a fixed  representation of quantum states, such as the density matrix, but builds its own state representation in the training phase, and (2) does not assume a fixed  representation of quantum  measurements, such as the POVM, but works with a generic parametrization.  In general, the parametrization can be provided in an encrypted form, which conceals certain properties of the quantum hardware, such as the dimensionality of the underlying quantum system.  
    }
    \label{fig:NetworkStructure}
\end{figure*}

%\end{widetext}

The structure of GQNQ is illustrated in   Fig.~\ref{fig:NetworkStructure},  where we also provide  a comparison with quantum state tomography. 
The first step is to produce a  representation of the unknown quantum state $\rho$.
%characterize   quantum states from the available data.   
     In GQNQ, this step is carried out  by  a representation network, which computes a function $f_{\bm{\xi}}$ depending on parameters $\bm \xi$ that are fixed after  the training phase (see Methods for details).  The representation network receives as input the parametrization of all measurements in $\cal S$ and their outcome statistics on the state  $\rho$ that needs to be characterized.     For each pair  $(\bm{m}_i  ,   \bm{p}_i )$, the representation network produces  a vector $\bm{r}_i  =  f_{\bm{\xi}}  (\bm{m}_i  ,   \bm{p}_i )$.  The  vectors corresponding to different pairs are then combined into a single vector $\bm{r}$ by an aggregate function $\mathcal{A}$.  For simplicity,  we take the aggregate function to be the average, namely $\bm{r}:=\frac{1}{s}\sum_{i=1}^s \bm{r}_i$. At this point,  the vector $\bm{r}$  is a representation of the quantum state $\rho$. 

While tomographic protocols strive to find the density matrix that  fits the measurement  data, GQNQ is not constrained to a specific choice of state representation.   This  additional freedom enables the network to construct lower-dimensional representations of quantum states with sufficiently regular structure, such as ground states in well-defined phases of matter, and to make predictions for states that did not appear in the training phase.  Notice also that the tomographic reconstruction of the density matrix using statistical estimators,  such as maximum likelihood and maximum entropy~\cite{teo2011}, is generally more time-consuming  than the evaluation of the function $f_{\bm{\xi}}$, due to the   computational complexity of the estimation procedure.

Once a state representation has been produced, the next  step is to predict the outcome statistics for a new quantum measurement on the state $\rho$.   In quantum tomography, the prediction is generated by applying the Born rule on the estimated density matrix.  In GQNQ, the task is achieved by  a generation network~\cite{eslami2018}, which computes a function  $g_{\bm{\eta}}$ depending on some  parameters $\bm{\eta}$ that are fixed after the training phase. The network   receives as input the state representation $\bm{r}$ and the parametrization $\bm{m}'$ of the desired measurement $\bm{M}'  \in  \mathcal M\setminus \mathcal S$, and produces as  output a vector  $\bm{p}'= g_{\bm{\eta}}(\bm{r},\bm{m}')$ that approximates the outcome statistics of the measurement $\bm{M}'$ on  the state $\rho$. 

Another difference with quantum tomography is that GQNQ does not require a specific representation of quantum measurements in terms of POVM operators. Instead,  a  measurement parametrization  is sufficient for GQNQ to make its predictions, and the parametrization can even  be provided in an encrypted form.  % The feature makes GQNQ is suitable for  applications where one party holds the quantum hardware and another party holds classical computational resources for data analysis.  
Since GQNQ  does not require the description of the devices to be provided in clear, it can be used to perform data analysis on a public server, without revealing properties of the quantum hardware, such as the dimension of the underlying quantum system.

So far, we described the GQNQ  procedure  for learning a single quantum  state $\rho$.  In the case of multiple states,  the same procedure is repeated on each state, every time choosing a (generally different)  set of measurements $\mathcal S$.    Crucially, the network does not need any parametrization   of the quantum states, neither it needs the states to be   sorted  into different classes.   For example, if the states correspond to different phases of matter, GQNQ does not need to be told which state belongs to which phase. This feature will be important for  the applications to state clustering and classification illustrated later in this paper.

The internal structure of the representation and generation networks is discussed  in Supplementary Note~\ref{sec:implementation}.  The  parameters $\bm{\xi}$ and $\bm{\eta}$ are determined in the training phase, in which 
%the representation network and the generation network are trained jointly to minimize the   difference between the predicted outcome distributions and the real distributions,  on average over all possible measurements (see Methods for more details).   To train GQNQ, one
GQNQ is provided with pairs $(\bm{m},  \bm{p})$ consisting of the measurement parametrization/measurement statistics for a fiducial set of measurements ${\cal M}_*  \subseteq \cal M$, performed on  a fiducial  set of quantum states $\cal Q_*$.     In the numerical experiments  provided in the Results section,  we choose ${\cal M}_*=\cal M$, that is, we provide the network with the statistics of all the measurement in $\cal M$. In the typical scenario,  the fiducial states and measurements are known, and the training can be done offline,  using computer simulated data rather than actual experimental data.

    We stress that  the parameters $\bm{\xi}$ and $\bm{\eta}$ depend only on  the fiducial sets $\cal M_*$ and $\cal Q_*$ and on the corresponding measurement data,  but do not depend   on the unknown quantum states  that will be characterized later,  nor on the subsets of measurements that will be performed on these states. Hence, the network does not need to be re-trained  when it is used to characterize a new quantum state $\rho$, nor  to be re-trained when one changes the subset of performed measurements $\cal S$.
    
    %For GQNQ to achieve a satisfactory performance, the fiducial states in the training set need to be sufficiently similar to the states on which the network will have to make predictions.  In this sense, the choice of a fiducial set can be regarded as a weak form of supervision. % Supervised machine learning algorithms were shown to offer advantages in  learning  the ground state properties of gapped Hamiltonians~\cite{huang2021} and in  a number of quantum problems \cite{huang2021power}.   Accordingly, it is natural to expect that a suitably trained GQNQ can   outperform unsupervised learning models for quantum state characterization.   {\color{blue}  [Does the paper lose something if we remove the last two sentences?  It is slightly awkward to say that we expect something without having any concrete result to show.]}
    %such as restricted Boltzmann machines. 

Summarizing, the main structural features of GQNQ are
\begin{itemize}

      \item  Offline, multi-purpose training: training can be done offline using computer generated data. Once the training has been concluded, the network can be used to characterize and compare  multiple  states.
     %(other than the information about   data correspond to which state.
   \item Measurement flexibility: %a trained GQNQ depends only on  the set of fiducial measurements and on the set of fiducial states, but not on the subset of measurements $  \cal S$ used to learn an unknown quantum state. 
  after the training has been completed, the experimenter can freely choose which subset of measurements $\cal S \subset \cal M$ is performed on the  unknown  quantum states.
   \item Learner-blindness: the parametrization of the measurements can be provided in an encrypted form.   No parametrization of the states is needed. 
\end{itemize}
Later in the paper, we will show that  GQNQ can be adapted  to an online version of the state learning task~\cite{aaronson2019online},  thus achieving the additional feature of 
\begin{itemize}
    \item Online prediction:    predictions can be updated as new measurement data become available.
\end{itemize}

\medskip{}{}{}{}

 {\bf Quantum state learning in spin  systems.}\label{subsec:numerical} A natural test bed for our neural network model is provided by quantum spin systems~\cite{schollwock2008,Sam13}.
In the following, we consider ground states of the one-dimensional transverse-field Ising model  and of the XXZ model,  both of which are significant  for many-body quantum simulations~\cite{friedenauer2008,kim2010,islam2011}.  These two models correspond to the Hamiltonians
\begin{equation}\label{transIsing}
    H=-\left(\sum_{i=0}^{L-2} J_i\sigma_i^z\sigma_{i+1}^z+\sum_{j=0}^{L-1} \sigma_j^x\right),
\end{equation}
and 
\begin{equation}\label{xxzmodel}
    H=-\left[\sum_{i=0}^{L-2}   \Delta_i(\sigma_i^x\sigma_{i+1}^x+ \sigma_i^y\sigma_{i+1}^y) + \sigma_i^z\sigma_{i+1}^z \right] \, ,  
\end{equation}
respectively.   In the Ising Hamiltonian (\ref{transIsing}), positive (negative) coupling parameters $J_i$ correspond to ferromagnetic (antiferromagnetic) interactions.  For the XXZ Hamiltonian (\ref{xxzmodel}), the ferromagnetic phase corresponds to coupling parameters $\Delta_i$ in the interval $(-1,1)$.   If instead the coupling parameters fall in the region $(-\infty,-1)\cup(1,\infty)$, the Hamiltonian is said to be  in the  XY phase~\cite{YY66}.

We start by considering a system of six qubits as example. For the ground states of the Ising model~(\ref{transIsing}),  we choose each coupling parameter $J_i$ at random  following a Gaussian  distribution with  standard deviation  $\sigma = 0.1$  and mean $J$.  For $J>0$ ($J<0$), this random procedure has a bias towards ferromagnetic (antiferromagnetic) interactions. For $J=0$, ferromagnetic and antiferromagnetic interactions are equally likely.
%and we call the corresponding ground states ``unbiased.''  
Similarly, for the ground states of the XXZ model~(\ref{xxzmodel}), we choose each parameter $\Delta_i$  at random  following a Gaussian distribution with standard deviation $0.1$ and mean value $\Delta$.  When   $\Delta$ is in the interval $(-1,1)$  ($(-\infty,-1)\cup(1,\infty)$), this random procedure has  a bias towards interactions of the ferromagnetic (XY) type.  

 In addition to the above  ground states, we also consider locally rotated GHZ states, of the form $\otimes_{i=1}^6 U_i\ket{\text{GHZ}}$ with $\ket{\text{GHZ}}=\frac{1}{\sqrt{2}}(\ket{000000}+\ket{111111})$ and locally rotated W states, of the form    $\otimes_{i=1}^6 U_i\ket{\text{W}}$ with  $\ket{\text{W}}=\frac{1}{\sqrt{6}}(\ket{100000}+\dots+\ket{000001})$, where  $   (U_i)_{i=1}^6 $ are unitary matrices of the form $U_i=\exp[-\text{i}\theta_{i,z} \sigma_{i,z}]\exp[-\text{i}\theta_{i,y}\sigma_{i,y}]\exp[-\text{i}\theta_{i,x}\sigma_{i,x}],$
where the angles  $\theta_{i,x}, \theta_{i,y}, \theta_{i,z}\in [0,\pi/10]$ are chosen independently and uniformly at random for every $i$.   

For the set of all possible measurements  $\cal M$, we chose the 729 six-qubit measurements consisting of local Pauli measurements on each qubit. 
 To parameterize the measurements in $\cal M$,  we provide the entries in the corresponding Pauli matrix at each qubit, arranging the entries in a 48-dimensional real vector.   
The dimension of state representation $\bm{r}$ is set to be
%Here, we set the dimension of $\bm r$ to be 
$32$, which is half of the Hilbert space dimension.  In Supplementary Note~\ref{sec:hyperparameter} we discuss how the choice of dimension of $\bm{r}$ and the other  parameters of the network affect the performance of GQNQ.

   GQNQ is trained using  measurement data from measurements in $\cal M$  on  states of the above four types (see Methods for a discussion of the data generation techniques).  We consider both the scenarios where all training data come from states of the same type, and where states of different types are used.   In the latter case, we do not provide the network with any label of the state type.  After training, we test GQNQ on   states of the four types described above.  To evaluate the performance of the network, we compute the classical fidelities between the predicted  probability distributions and the correct distributions computed from the true states and measurements.        For each test state,  the classical fidelity is averaged  over all possible  measurements in $\mathcal M\setminus\mathcal{S}$, where  $\mathcal{S}$ is a random subset of $30$ Pauli measurements.  Then, we average the fidelity over all possible test states.

The results  are summarized in Table~\ref{tab:sixqubit}.  Each row shows the  performances of one particular trained GQNQ when tested using the measurement data from (i) $150$ ground states of Ising model with  $J\in \{ 0.1, \dots, 1.5\}$, (ii)  $150$ ground states of Ising model with  $J\in \{-1.5, -1.4, \dots, -0.1\}$, where $10$ test states are generated per value of $J$, (iii)  $10$  ground states of Ising model with $J=0$,  (iv) $190$ ground states of XXZ model with $\Delta\in \{-0.9, -0.8, \dots, 0.9\}$, (v) $100$ ground states of XXZ model with $\Delta\in\{-1.5, -1.4, \dots, -1.1\}\cup \{1.1, 1.2, \dots, 1.5\}$, where $10$ test states are generated per value of $\Delta$, (vi) all the states from (i) to (v), (vii) $200$ locally rotated GHZ states     (viii) $200$ locally rotated W states   (vii),   (ix) all the states from (i) to (v), together with (vii) and (viii). 
%{\color{cyan}[Add that each row is a trained model]}
     In the second column, the input data given to GQNQ is the  true probability distribution computed with the Born rule, while in the third and fourth columns, the input data given to GQNQ during test is the finite statistics obtained by sampling the true outcome probability distribution $50$ times and $10$ times, respectively.

\begin{widetext}

\begin{table}[h!]
    \centering
    \caption{Average classical fidelities between the predictions of GQNQs and the ground truths with respect to different types of six-qubit states.
    }
    \begin{tabular}{c|c|c|c}
    \hline\hline
      Types of states for  training and test  & noiseless & $50$ shots & $10$ shots \\ \hline
      (i) Ising ground states with ferromagnetic bias      &   0.9870 & 0.9869 &  0.9862\\
      (ii) Ising  ground states with antiferromagnetic bias   &   0.9869 &  0.9867  & 0.9849 \\
      (iii) Ising ground states with no bias  & 0.9895 & 0.9894 & 0.9894 \\
      (iv) XXZ ground states with ferromagnetic bias   & 0.9809 & 0.9802 &  0.9787 \\
      (v) XXZ ground states with XY phase bias    &  0.9601 & 0.9548 & 0.9516 \\
      (vi) (i)-(v)  together   &  0.9567 &  0.9547 & 0.9429 \\
      (vii) GHZ state with local rotations &  0.9744 & 0.9744 & 0.9742 \\
      (viii) W state with local rotations &  0.9828 & 0.9826 & 0.9821   \\
      (ix) (i)-(v), (vii) and (viii) together   &   0.9561 & 0.9543 &  0.9402 \\
    \hline\hline
    \end{tabular}
    
    \label{tab:sixqubit}
\end{table}

\begin{figure}[ht]
    \centering
    \includegraphics[width=0.8\textwidth]{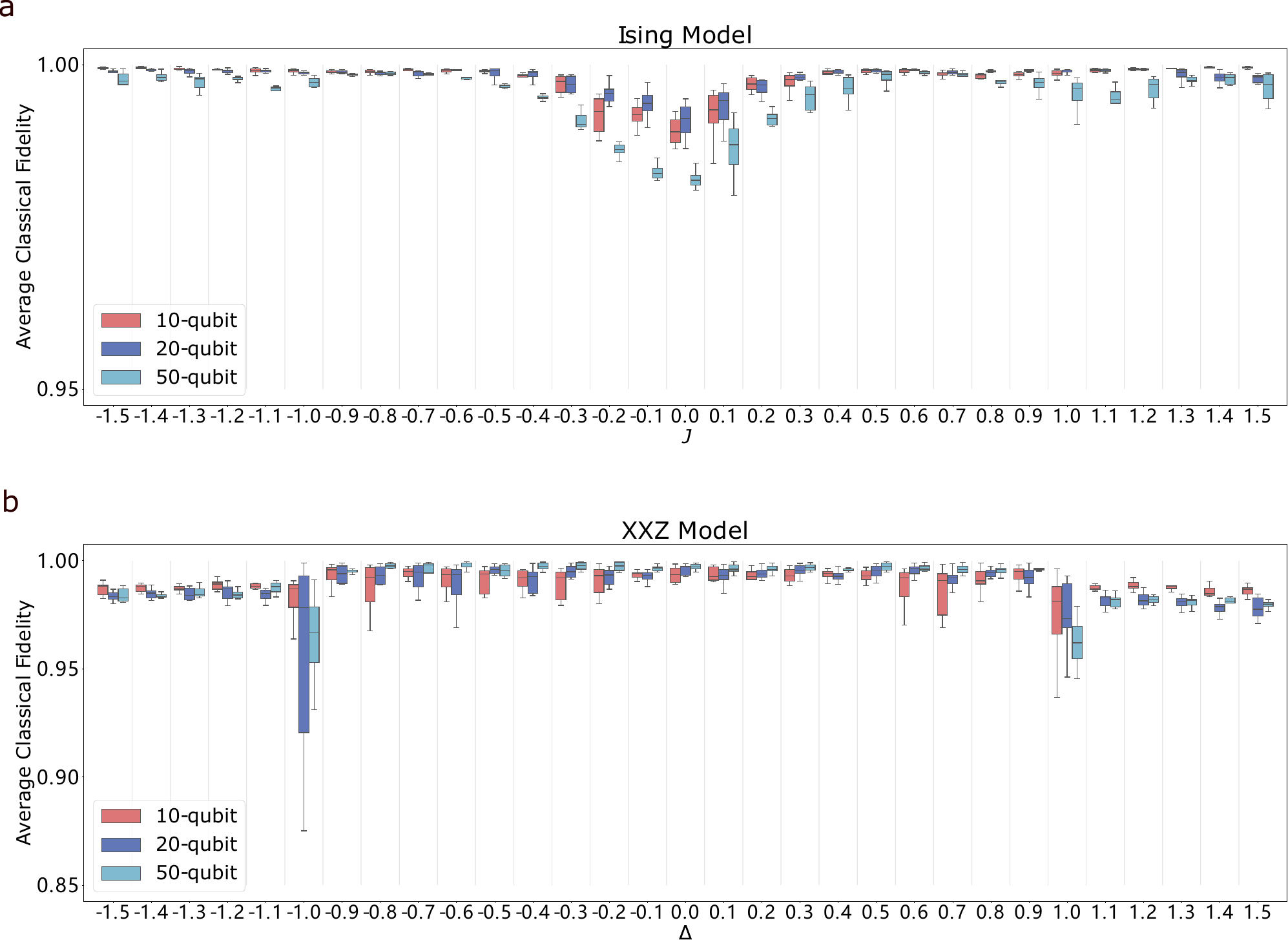}
    \caption{
    Performances of GQNQs on Ising model ground states and XXZ model ground states visualized by boxplots \cite{williamson1989box}. 
    Figure (a) shows the average classical fidelities of predictions given by three GQNQs for ten-, twenty- and fifty-qubit ground states of Ising model~(\ref{transIsing}), respectively, with respect to different values of $J\in \{-1.5, -1.4, \dots, 1.5\}$.
    Figure (b) shows the performances of another three GQNQs for ten-, twenty- or fifty-qubit ground states of XXZ model~(\ref{xxzmodel}), respectively, with respect to different values of $\Delta\in\{-1.5, -1.4, \dots, 1.5\}$.
   % In both cases, the class of measurements $\mathcal M$ is the set of all nearest-neighbour two-qubit Pauli-basis measurements and $\mathcal S$ is a subset containing $s=30$ measurements randomly chosen from $\mathcal{M}$. 
   Given outcome probability distributions for all $\bm{m}\in\mathcal S$, each box shows the average classical fidelities of predicted outcome probabilities, averaged over all  measurements in $\mathcal{M} \setminus \mathcal S$, for ten instances. }
    \label{fig:groundState}
\end{figure}
  
\end{widetext}
  
The results shown in Table \ref{tab:sixqubit} indicate  that the performance with finite statistics is only slightly lower than the performance in the ideal case.
% {\color{cyan}with the largest discrepancies appearing at phase transition points.}   
It is also worth noting that GQNQ maintains a high fidelity even when used on multiple types of states.     

Recall that the results in Table \ref{tab:sixqubit} refer to the scenario where GQNQ is trained with the full set of six-qubit Pauli measurements, which is informationally complete. An interesting question is whether the learning performance would still be good if the training used a non-informationally complete set of measurements. In Supplementary Note~\ref{sec:generalization}, we show that fairly accurate predictions can be made even if  $\cal M$ consists only  of $72$ randomly chosen Pauli measurements. 

While GQNQ makes accurate predictions for state families with sufficient structure,  it should not be expected to work universally well on all possible quantum states.  In Supplementary Note~\ref{sec:randomState}, we considered the case where the network is  trained and tested   on arbitrary six-qubit states, finding that the performance of GQNQ  drops drastically.   
%The situation is different when the set of possible measurement is restricted. For example, if  $\cal M$ consists of  $72$ random chosen six-qubit Pauli basis measurements and $\mathcal S$ consists of  $30$ random  measurements in $\cal M$, we find that the performance still remains {\color{red} pretty good} (see Supplementary Note $3$ for details). 
In Supplementary Note~\ref{sec:overfitting}, we  also provide numerical experiments on  the scenario where some types of states are overrepresented in the training phase, potentially causing overfitting when GQNQ is used to characterize unknown states of an underrepresented type.

%(in particular, we choose this type of the tested state to appear 10 times less frequently than the other types of states). The results show that the performance of GQNQ with unbalanced training data depends on the state under consideration. For W states with local rotations we find that unbalanced training data has little effect on the performance. The situation is similar for the ground states of the Ising model in the ferromagnetic phase. In contrast, the prediction for XXZ model in the XY phase drops to $0.73$ when the  training data are unbalanced.  

%We also consider the case where the network is required to work on arbitrary six-qubit states. In this case, the performance of GQNQ  drops drastically, and    becomes  comparable to  simply guessing the uniform distribution. This drop in performance is likely to be  due to the lack of well-structured pattern in the data, which makes it hard for the network to learn.  

We now  consider multiqubit states with $10$, $20$, and $50$ qubits, choosing the measurement set   $\cal M$  to consist of all  two-qubit Pauli measurements on  nearest-neighbor qubits and $\mathcal S$ a subset containing $s=30$ measurements randomly chosen from $\mathcal{M}$. Here the dimension of state representation $\bm{r}$ is chosen to be $24$, which guarantees a good performance in our numerical experiments.

For the Ising model, we choose  the coupling between each nearest-neighbour pair of spins to be either consistently ferromagnetic for $J \ge 0$ or consistently antiferromagnetic for $J<0$: for $J\ge 0$ we replace each coupling $J_i$ in Eq.\ (\ref{transIsing}) by $|J_i|$, and for $J<0$ we replace $J_i$ by $-|J_i|$.  
% For $J=0$, instead, we allow both ferromagnetic and antiferromagnetic couplings.
The results are illustrated in Fig. \ref{fig:groundState}.   The figure shows that the average classical fidelities in both ferromagnetic and antiferromagnetic regions are close to one, with  small  drops around the phase transition point $J=0$. 
The case where both ferromagnetic and anti-ferromagnetic interactions are present  is studied in Supplementary Note~\ref{sec:additional},  where we observe that the learning performance is less satisfactory in this scenario.

For XXZ model, the average classical fidelities in the XY phase are lower than those in the ferromagnetic interaction region, which is reasonable due to higher quantum fluctuations in the XY phase~\cite{Sam13}.  At the phase transition points $\Delta=\pm 1$,  the average classical fidelities drop more significantly, partly  because the abrupt changes of ground state properties at the critical points make the quantum state less predictable, and partly  because the states at phase transition points are less represented in the training data set.

\medskip 

{\bf Quantum state learning on a harmonic oscillator.}  We now test GQNQ on states encoded in harmonic oscillators, i.e. continuous-variable quantum states, including single-mode Gaussian states, as well as non-Gaussian states such as cat states and GKP states~\cite{gottesman2001},  both of which are  important for fault-tolerant quantum computing~\cite{gottesman2001,albert2018}.
For the  measurement set $\cal M$, we choose   $300$ homodyne measurements, that is, $300$  projective measurements associated to quadrature operators of the form $ ( e^{  i\theta}  \,\hat{a}^\dag +e^{-i\theta}\,  \hat{a})/2$, where $\hat{a}^\dag$ and $\hat{a}$ are bosonic creation and annihilation operators, respectively, and  $\theta$ is a uniformly distributed phase in the interval $[0,\pi)$.  
 For the subset $\cal S$, we pick  $10$ random quadratures.  
For the parametization of the measurements, we simply choose the corresponding phase $\theta$. 
Since the homodyne measurements have an unbounded and continuous set of outcomes, here we truncate the outcomes into a finite interval  (specifically,  at $\pm 6$) and discretize them, dividing the interval into  $100$ bins of equal width. % {\color{blue} what is the dimension of the vector $\bm{r}?$}
The dimension of the representation vector $\bm{r}$ is chosen to be $16$.

%\begin{widetext}

\begin{table*}[t]
    \centering
        \caption{  Performances of GQNQ on continuous-variable quantum states. 
    }
    \begin{tabular}{c | c |c |c|c|c|c}
    \hline\hline
       Type of states for training and test  & i.\ noiseless & worst case for i  & ii.\ $\sigma$(noise)$=0.05$ & worst case for ii & iii.\ $\sigma(\theta)=0.05$ & worst case for iii \\
        \hline
        (i) Squeezed thermal states & $0.9973 $ & $0.9890$  & $0.9964$ & $0.9870$ & $0.9972$ & $0.9889$  \\
        (ii) Cat states & $0.9827 $ & $0.9512$  & $0.9674 $ & $0.9053$ & $0.9822$ & $0.9461$ \\
        % (1) and (2) together & $0.9878 $ & $0.9604$  & $0.9753$ & $0.9391$ & $0.9874$ & $0.9575$  \\
        (iii) GKP states & $0.9762 $ & $0.9405$  & $0.9746 $ & $0.9359$  & $0.9758$ &  $0.9405$\\
        (iv) (i)-(iii) together  & $0.9658$ & $0.9077$  & $0.9264$ & $0.8387$ & $0.9643$ & $0.9030$ \\
        \hline\hline
    \end{tabular}
    \label{tab:continuous-outcome}
\end{table*}

%online learning figure
\begin{figure*}
    \centering
    \includegraphics[width=0.65\textwidth]{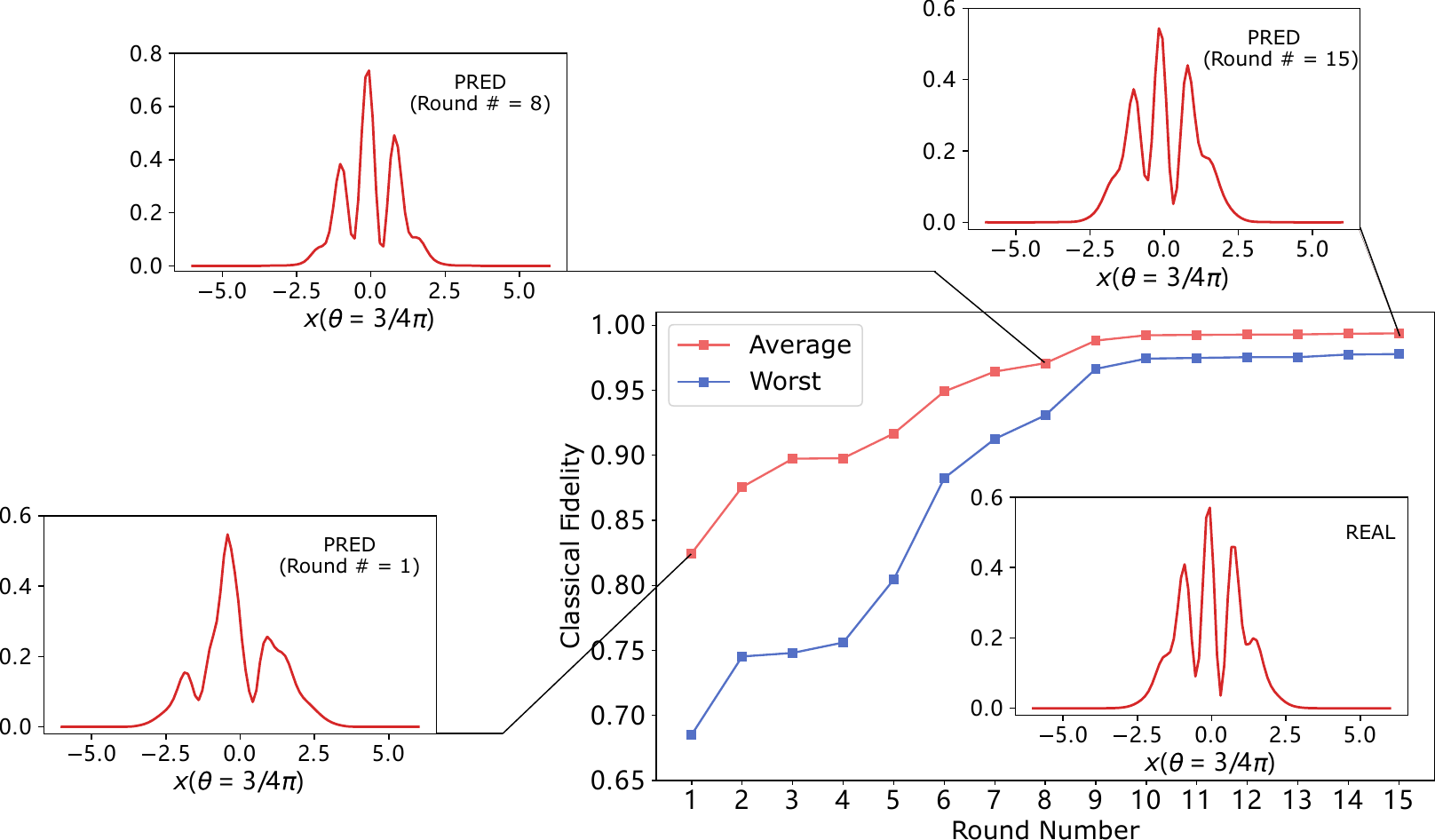}
    \caption{Online learning of cat states in 15 time steps. 
    Each red point shows the  classical fidelity, averaged over all the possible measurements in $\cal M$ and over all the cat states in the test set, which we take to be the same as in the experiments in the previous section.
    Each blue point is the worst-case classical fidelity over all possible query measurements, averaged over all the test states. Real outcome statistics and predicted outcome statistics at quadrature phase $\theta=3/4\pi$ for an example cat state $\ket{2.22+1.41\text{i}, \pi/4}_{\text{cat}}$ are plotted.  }
    \label{fig:onlineLearning}
\end{figure*}

%\end{widetext}

In Table~\ref{tab:continuous-outcome} we illustrate the performance of GQNQ on 
(i) $200$ squeezed thermal states
%$\rho_{V,s}^{\text{G}}:=\text{e}^{\frac{1}{2}(s^* \hat{a}^2- s\hat{a}^{\dagger\, 2})}\rho_{V}\text{e}^{-\frac{1}{2}(s^* \hat{a}^2- s\hat{a}^{\dagger\, 2})}$
%where $\rho_V$ is a thermal state 
with thermal variance $V\in [1,2]$ and squeezing parameter $s$ satisfying $|s|\in [0,0.5]$, $\text{arg}(s)\in [0, \pi]$, (ii) $200$  cat states corresponding to superpositions of coherent states with opposite amplitudes
$
\ket{\alpha,\phi}_{\text{cat}}:=\frac{1}{\sqrt{\mathcal N}}(\ket{\alpha}+\text{e}^{\text{i}\phi}\ket{-\alpha})
$,
where $\mathcal N=2(1+\text{e}^{-|\alpha|^2}\cos\phi)$, $|\alpha|\in [1,3]$ and $\phi\in \{0, \frac{\pi}{8}, \dots, \pi\}$,    (iii) $200$ GKP states that are superpositions of displaced squeezed states
$
\ket{\epsilon, \theta,\phi}_{\text{gkp}}:=\text{e}^{-\epsilon\hat{n}}\left(\cos\theta \ket{0}_{\text{gkp}}+\text{e}^{\text{i}\phi}\sin\theta\ket{1}_{\text{gkp}}\right)
$
where $\hat{n}=\hat{a}^\dagger\hat{a}$ is the photon number operator, $\epsilon\in [0.05, 0.2]$, $\theta\in [0, 2\pi)$, $\phi\in [0,\pi]$, and $\ket{0}_{\text{gkp}}$ and $\ket{1}_{\text{gkp}}$ are ideal GKP states, and (iv) all the states from (i), (ii), and (iii).    

For each type of states, we provide the network with  measurement data from $s=10$ random homodyne measurements, considering  both the case where the  data is noiseless   and the case where it is noisy.   The noiseless case is shown in the second and third columns of Table~\ref{tab:continuous-outcome}, which show the classical fidelity in the average and worst-case scenario, respectively.  In the noisy case, we consider both noise due to finite statistics, and noise due to an inexact specification of the measurements in the test set. 
The effects  of finite statistics are modelled by  adding  Gaussian noise to  each  of the outcome probabilities of the measurements in the test.  The inexact specification of the test measurements   is modelled by rotating each quadrature by a random angle~$\theta_i$, chosen independently for each measurement according to a Gaussian distribution.   
The fourth and the fifth columns of Table~\ref{tab:continuous-outcome} illustrate the effects of finite statistics, showing the classical fidelities in the presence of  Gaussian added noise with variance $0.05$.
    In the sixth and seventh columns, we include the effect of an inexact specification of the homodyne measurements, introducing Gaussian noise with  variance $0.05$.
    In all  cases, the classical fidelity of predictions are computed  with respect to the ideal noiseless probability distributions.

    In Supplementary Note~\ref{sec:additional} we also provide a more detailed comparison  between the predictions and the corresponding ground truths in terms of actual probability distributions, instead of their classical fidelities.  
%  {\color{red} It is also worth mentioning that the set of measurements $\mathcal M$ can also be chosen to be a small set.  For instance, when $\mathcal M$ consists of only $10$ homodyne measurements and $\mathcal S$ is a subset of five homodyne measurements, the generalization performance of GQNQ remains good (see Supplementary Note 3 for details).}{\color{blue}  I suggest to remove the above part in red. It does not give any further insight to the reader, and it may even raise objections: like, "true that the set $\cal M$ is smaller, but in our experiment the ratio between the size of $\cal M$ and $\cal S$ is more favourable"}. 

\iffalse
In the case of cat states, we also provide the full probability distributions predicted by GQNQ for two quadratures with 10 random quadratures in the training set, and we compare the predicted probability distributions with  exact ones.   The comparison is shown in  Fig.~\ref{fig:cat}.   {\color{blue}  [Do we need this figure? Of maybe we can move it to the Methods section, or to the Supplemental Material?]}
\fi

\medskip 

{\bf Application to online learning.}   After GQNQ has been trained, it can be used for the task of  online quantum state learning~\cite{aaronson2019online}.  In this task, the various pieces of data are provided to the learner at  different time steps. At the $i$-th time step, with $i\in  \{1,\dots,  n\}$,   the experimenter performs a measurement $\bm{M}_i$, obtaining the outcome statistics $\bm{p}_i$.  The pair $  (  \bm{m_i},  \bm{p_i})$ is then provided to the learner, who is asked to predict the measurement outcome probabilities for all   measurements in the set $\mathcal M \setminus  \mathcal S_i$ with  ${\cal S}_i:= \{\bm{M}_j\}_{j\le i}$.

Online learning with GQNQ can be achieved with the following procedure.   Initially, the state representation vector is set to  $\bm{r}  (0)=  (0,\dots,  0)$.  At the $i$-th time step,  GQNQ  computes the vector  $\bm{r}_i  =  f_{\bm{\xi}}  (\bm{m}_i, \bm{p}_i)$ and updates the state representation to $  \bm{r} (i)  =  [(i-1) \, \bm{r} (i-1)   +  \bm{r}_i]/i$. The updated state representation is then fed into the generation network, which produces  the required predictions. Note that updating the state representation does not require time-consuming operations, such as a maximum likelihood analysis.  It is also worth noting that   GQNQ does not need to  store all the measurement data received in the past: it only needs to store the state representation $\bm{r}(i)$ from one step to the next. 

A numerical experiment on online learning of cat states is provided in Fig.~\ref{fig:onlineLearning}. 
The figure shows the average classical fidelity at 15 subsequent time steps corresponding to 15 different homodyne measurements performed on copies of unknown cat states. The fidelity increases over time, confirming the intuitive expectation that the  learning performance should improve when more measurement data are provided.

%\begin{widetext}

\begin{figure*}
    \begin{center}
    \includegraphics[width=0.85\textwidth]{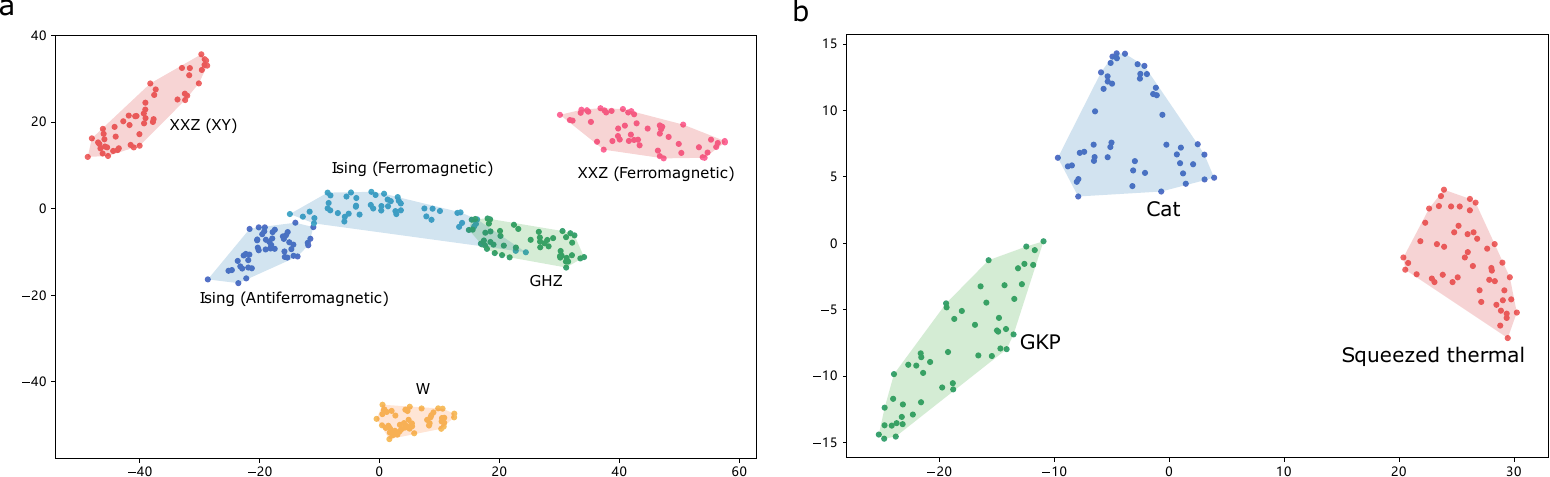}
    \end{center}
    \caption{Two-dimensional embeddings  of multiqubit and continuous-variable states. %Quantum state  clustering of multiqubit and continuous-variable states. 
    Subfigure (a) shows two-dimensional embeddings  of state representation vectors produced by GQNQ  on Ising model (ferromagnetic and antiferromagnetic) ground states, XXZ model (ferromagnetic and XY phase) ground states, locally rotated GHZ and W states.   Subfigure (b) shows two-dimensional embeddings 
    of the state representation vectors of squeezed thermal states, cat states and GKP states.   In both subfigures,  shaded areas are added to help visualize the various type of states. Note that the representation vectors  generated by GQNQ of states of the same type are near to each other in the two-dimensional embeddings.}
    \label{fig:clustering}
\end{figure*}

%\end{widetext}

 \medskip 
 
{\bf Application to state clustering and classification.}
The state representation  constructed by  GQNQ can also be used to perform  tasks other than predicting the  outcome statistics of  unmeasured POVMs.  One such task is  state clustering, where  the goal is to group the  representations of different quantum states into multiple disjoint sets in such a way that  quantum states of the same type fall into the same set. 

We now show that clusters naturally emerge from the state representations produced by GQNQ.  To visualize the clusters,  we  feed the state representation vectors into a $t$-distributed stochastic neighbor embedding ($t$-SNE) algorithm~\cite{van2008visualizing}, which produces a mapping of the representation vectors into a two-dimensional plane, according to their  similarities.  We performed numerical experiments using  the types  of six-qubit   states in Table~\ref{tab:sixqubit} and the  types of continuous-variable states in Table~\ref{tab:continuous-outcome}.  For simplicity, we restricted the analysis to  state representation vectors constructed from noiseless input data.  

The results of our experiments are shown in  Fig.~\ref{fig:clustering}.  The figure shows that  states with significantly different physical properties correspond to distant points in the two-dimensional embedding, while states with similar properties naturally appear in  clusters.  
For example, the ground states of the ferromagnetic XXZ model and the ground states in the gapless XY phase are clearly separated in Fig.~\ref{fig:clustering}(a), in agreement with the fact that  there is an abrupt change of quantum properties  at the phase transition point.   On the other hand, in Fig.~\ref{fig:clustering}(a),
 the ferromagnetic region of the Ising model is next to the antiferromagnetic region, 
both of which are gapped and short-range correlated. 
% {\color{blue}  [Is there any physical explanation for this similarity?  It is quite striking that different phases in the XXZ model give far away clusters, while different phases in the Ising model appear basically in the same cluster... I was  expecting ferromagnetic and anti-ferromagnetic states to be very different from each other, why is it that they overlap? ]} 
 The ferromagnetic region of the Ising model appears to have some overlap with the region of GHZ states with local rotations, in agreement with the fact that the GHZ state is approximately a ground state of the ferromagnetic Ising model in the large  $J$ limit. 

The visible clusters in the two-dimensional embedding suggest  that any unsupervised clustering algorithm could effectively cluster the states according to their representation vectors. To confirm this intuition,  we applied a Gaussian mixture model \cite{bishop2006pattern} to the state representation vectors  and chose the number of clusters to be equal to the actual number of state types (six for the six-qubit states, and three for the continuous-variable states). The portion of states whose types match the clusters is $94.67\%$ for the six-qubit states, and $100\%$ for the continuous-variable states.

The state representation produced by GQNQ can also  be used to predict physical properties in a supervised model where an additional neural network is provided with labelled examples of states with a given property.  In this setting, supervision can enable a more refined classification of quantum states, compared to the unsupervised clustering discussed before. 

To illustrate the idea, we considered the problem of distinguishing between two different regimes in the Ising model, namely a regime where ferromagnetic interactions dominate ($J>1$), and a regime both ferromagnetic and antiferromagnetic interactions are present ($0<J<1$).  For convenience, we refer to these two regimes as to the pure and mixed ferromagnetic regimes, respectively.  We use an additional neural network to learn whether a ground state corresponds to a Hamiltonian in the pure ferromagnetic regime or in the mixed one, using the state representation $\bm{r}$ of Ising ground states with ferromagnetic bias obtained from noiseless measurement data.
 The prediction reaches a success rate of $100\%$, $100\%$ and $99\%$ for ten-qubit, twenty-qubit and fifty-qubit ground states in our test sets, respectively.  These high values can be contrasted with the clustering results in Fig. 
\ref{fig:clustering}, where the pure ferromagnetic regime and the mixed one  appear close to each other in the two-dimensional embedding.

%\subsection*{Learning with Prior Knowledge of POVM Measurements}

%In our model above, both the number $n$ of POVM measurements and each POVM measurement $M_i$ in $\mathcal M$ are arbitrary, such that our model can be used in various scenarios like adversarial online learning.
%However, there is a tradeoff between generality of our model and its performance.
%If we have prior knowledge about the set $\mathcal M$, then we can upgrade the structure of GQNQ and train the representation network to enhance its capability to learn features of any quantum state only from the measurement data with respect to POVM measurements in $\mathcal M$.

\section{Discussion}

%To our knowledge, no previous approach possesses all these features. 
%For example, the data-driven approach in Ref.~\cite{Ahmed2021PRL} assumes the same  set of measurements both for the training and for the state learning phases, with the consequence that the network needs to be retrained if one wants to change the measurement set $\cal S$.   

%{\color{blue}  [Actually, there is a problem with this intuition: the input of our generative network is the vector $\bm{r}$, not the measurement data.  Any comment?  ]}{\color{cyan}[Ge: The generative query network includes both the representation network and the generative network, so the input of our generative query network is the measurement data.]}

%\begin{widetext}

\begin{table*}[!htbp]
\centering
\caption{Performances of GQNQ on cat states as an unsupervised learner}
\begin{tabular}{c|c|c|c|c}
\hline\hline
state & %$s=100$ (Avg)  & $s=100$ (Worst) & 
$s=50$ (Avg) & $s=50$ (Worst) & $s=10$ (Avg) & $s=10$ (Worst) \\\hline
$\ket{2, 0}_{\text{cat}}$ & %$0.9918$ & $0.9627$ & 
$0.9918$ & $0.9614$ & $0.9912$ & $0.9610$ \\
$\ket{2, \pi/4}_{\text{cat}}$ & %$0.9918$ & $0.9625$ & 
$0.9917$ & $0.9602$ & $0.9745$ & $0.9236$ \\
$\ket{2.22 + 1.41\text{i}, \pi/4}_{\text{cat}}$ &  % $0.9790$ & $0.9253$ &
$0.9779$ & $0.9171$ & $0.9671$ & $0.9133$ \\
\hline\hline
\end{tabular}
\label{tab:singleState}
\end{table*}

%\end{widetext}

Many works have explored the use of generative models for quantum state characterization~\cite{torlai2018np,Torlai2018prl,carrasquilla2019,tiunov2020,Ahmed2021PRL}, and an approach based on representation learning was recently proposed by Iten {\em et al}~\cite{Iten2020}. 
%A key  structural difference between GQNQ and these previous approaches is  that, rather than feeding the raw measurement data directly into a neural network that makes predictions, GQNQ uses a representation network to build its own description of quantum states.  This description is then ``interpreted'' by a generation network that makes predictions about the statistics of new quantum measurements.    
 % The division between representation and generation network offers new features at the training level.    
  The key difference between GQNQ and previous approaches concerns the training phase.  In most previous works,  the neural network is trained to reconstruct a single quantum state from experimental data.   While this procedure can in principle be applied to learn any  state,  the training is state-specific,  and the information learnt by the network through training on a given state cannot be automatically transferred to the reconstruction of a different quantum state, even if that state is of the same type.  In contrast, the training of GQNQ works for multiple quantum states  and for states  of multiple types, thus  enabling a variety of tasks, such as quantum state clustering and classification.

  Another  difference  with previous works is that  the training phase for GQNQ can use classically simulated data, rather than  actual experimental data.   In other words, the training can be carried out in an  offline mode, before the quantum states that need to be characterized become available.  By moving the training to offline mode,   GQNQ can be  significantly faster than  other data-driven approaches that need to be trained with experimental data from unknown quantum states.      The flip side of this advantage, however, is that offline training requires a partial supervision, which is not required in other state reconstruction approaches ~\cite{torlai2018np,Torlai2018prl,carrasquilla2019}.  Indeed, the  training of GQNQ requires  quantum states in the same family as the tested state, and in order to implement the training offline one needs a good guess for the type of quantum state that will need to be characterized.

  The situation is different if the training is done online, with actual experimental data provided from the quantum state to be characterized.   In this setting, GQNQ behaves as  a completely unsupervised learner  that predicts the outcome statistics of  unperformed measurements using measurement data obtained solely from the quantum state under consideration. 
  Note that in this case the set of fiducial measurements $\mathcal M_*$ coincides with the set of performed measurements ${\cal S}  \subset \cal M$. 
  The details of the training procedure are provided in Supplementary Note~\ref{sec:singleState}.    We performed numerical experiments in which GQNQ  was trained with data from a single cat state, using data from $10$ or $50$ homodyne measurements.  After the training, GQNQ was asked to predict the outcome statistics of a new randomly chosen homodyne  measurement. The results are summarized in Table~\ref{tab:singleState}, where we show both the average classical fidelities averaged over all query measurements and worst-case classical fidelities over all query measurements.

Finally, we point out that our learning model  shares some conceptual similarity with Aaronson's ``pretty good tomography''~\cite{aaronson2007}, which aims at producing 
  a hypothesis state that accurately predicts the outcome probabilities of measurements in a given set.  While in pretty good tomography the  hypothesis state is a density matrix,  the form of the state representation in GQNQ is determined by  the network itself.  
  The flexibility in the choice of state representation  allows GQNQ to find more compact descriptions  for sufficiently regular sets of states. On the other hand,  pretty good tomography is in principle guaranteed to work accurately for arbitrary quantum states, whereas the performance of GQNQ  can be more or less accurate depending  on the set  of states,  as indicated by our numerical experiments.  An important direction of future research is to find criteria to determine {\em a priori} which quantum state families can be learnt effectively by GQNQ. This problem is expected to be challenging, as similar criteria are still lacking even in the original application of generative query networks to classical image processing.

\section{Methods}

% \subsection*{Data Generation}
% In this subsection, we introduce our training/test dataset generation procedures for multi-qubit states and continuous-variable states.

%\subsection*{Neural network training}
%In this subsection, we discuss the training of our proposed GQNQ model for the state-learner. 

% {\color{red} a paragraph about training size, number of parameters}

{\em Data generation procedures.}   Here we discuss  the  training/test dataset generation procedures. In the numerical experiments for ground states of Ising models and XXZ models, the training set is composed of $40$ different states for each value of $J$ and $\Delta$, while the test set is composed of $10$ different states for each value of $J$ and $\Delta$. For GHZ and W states with local rotations, we generate $800$ states for training and $200$ states for testing. 

In the continuous-variable experiments,  we randomly generate $10000$ different states for each of the three families of squeezed thermal states, cat states, and GKP states.  We then split  the generated states  into a training set and testing set, with a ratio of  $4:1$.

In the testing stage, the noiseless probability distributions   for one-dimensional Ising models and XXZ models  are generated by  solving the ground state problem, either exactly (in the six qubit case) or approximately by density-matrix renormalization group (DMRG) algorithm~\cite{schollwock2005density} (for $10$, $20$ and $50$ qubits). The data of continuous-variable quantum states are generated by simulation tools provided by Strawberry Fields \cite{killoran2019strawberry}. % {\color{blue}  [What simulation?  Isn't everything analytical for these states?]}

\medskip 

{\em Network training.} The training data set of GQNQ  includes measurement data from $N$ quantum states, divided into $N/B$ batches of $B$ states each. For each state in a batch, a subset of measurements $\mathcal{M}_1 \subset \mathcal{M}$ is randomly picked, and the network is provided with all the pairs  $(\bm{m},  \bm{p})$, where $\bm m$ is the parametrization of a measurement in ${\cal M}_1$ and $\bm p$ is the corresponding vector of outcome probabilities on the state under consideration.   The network is then asked to make predictions on the outcome probabilities of the rest of the measurements in $\mathcal{M}\setminus \mathcal{M}_1$, and the loss is computed from the difference between the real outcome probabilities (computed with the Born rule) and the model's predictions (see Supplementary Note~\ref{sec:implementation} for the specific expression of the loss function). 
%The loss function is defined as the sum of losses over a batch of states. 
For each batch, we optimize the parameters $\bm{\xi}$ and $\bm{\eta}$ of GQNQ by updating them along the opposite direction of the gradient of the loss function with respect to $\bm{\xi}$ and $\bm{\eta}$,  using   Adam optimizer \cite{kingma2014adam} and batch gradient descent.     The pseudocode for  the training algorithm is also  provided in Supplementary Note~\ref{sec:implementation}.

The training is repeated for $E$ epochs. In each epoch of the training phase, we iterate the above procedure over the $N/B$ batches of training data.   For the numerical experiments in this  paper, we typically choose $B=30$ and $E=200$.   

\medskip  

{\em Network testing.}  After training, the parameters of GQNQ are fixed, and the performance is then tested with  these fixed parameter values.   
For each test state, we randomly select a subset $\mathcal S$ from the set $\mathcal M$ of POVM measurements, input the associated measurement data to the trained network, and ask it to predict the  outcome probabilities for all the measurements in $\mathcal M\setminus \mathcal S$. Then we calculate the classical fidelity between each output prediction and the corresponding ground truth.    

%{\em State representation dimension.} 
%{\color{red} The dimensions of the state representations are set to 
%Here, we set the dimension of $\bm r$ to be 
%$32$ for the scenario of six-qubit states, $24$ for the scenario of $10$-, $20$- and $50$-qubit states,} and $16$ for continuous-variable quantum states, all of which are much smaller than the degrees of freedom in their density matrices.

{\em Hardware.}
Our neural networks are implemented by the pytorch \cite{paszke2019pytorch} framework and trained on four NVIDIA GeForce GTX 1080 Ti GPUs.

% {\color{red} Here we randomly generate $10000$ different cat states, gaussian states and GKP states and split them into training and test sets with $80//20$ ratio.}

% {\color{red} In the experiments for ground states of Ising models and XXZ models, the training set is composed of $50$ different states for each $J$ or $\Delta$ while the test set is composed of $10$ different states for each $J$ or $\Delta$. As for the experiments for GHZ state with local rotation and W state with local rotation, we generate $1000$ states for training and $200$ states for test. }

%The implementation of the neural network model and the data sets we used shown in this manuscript are available at ...

\bibliography{refs.bib}

\section{acknowledgement}
This work was supported by funding from the Hong Kong Research Grant Council through grants no.\ 17300918 and no.\ 17307520, through the Senior Research Fellowship Scheme SRFS2021-7S02, the Croucher Foundation, and by  the John Templeton Foundation through grant 61466, The Quantum Information Structure of Spacetime (qiss.fr). 
YXW acknowledges funding from the National Natural Science Foundation of China through grants no.\ 61872318. Research at the Perimeter Institute is supported by the Government of Canada through the Department of Innovation, Science and Economic Development Canada and by the Province of Ontario through the Ministry of Research, Innovation and Science. The opinions expressed in this publication are those of the authors and do not necessarily reflect the views of the John Templeton Foundation.

\begin{widetext}
\section*{\uppercase{ Supplementary Notes}}
% \section{Appendix}
\section{Implementation details of GQNQ}
\label{sec:implementation}

\subsection{Structure of GQNQ}
As shown in Fig.~\ref{fig:gqnq_struc}, our proposed Generative Query Network for quantum state learning (GQNQ) is mainly composed of a representation network $f_{\bm{\xi}}$, an aggregate function $\mathcal{A}$ and a generation network $g_{\bm{\eta}}$. 

\begin{figure}[h]
    \centering
    \includegraphics[width=0.65\textwidth]{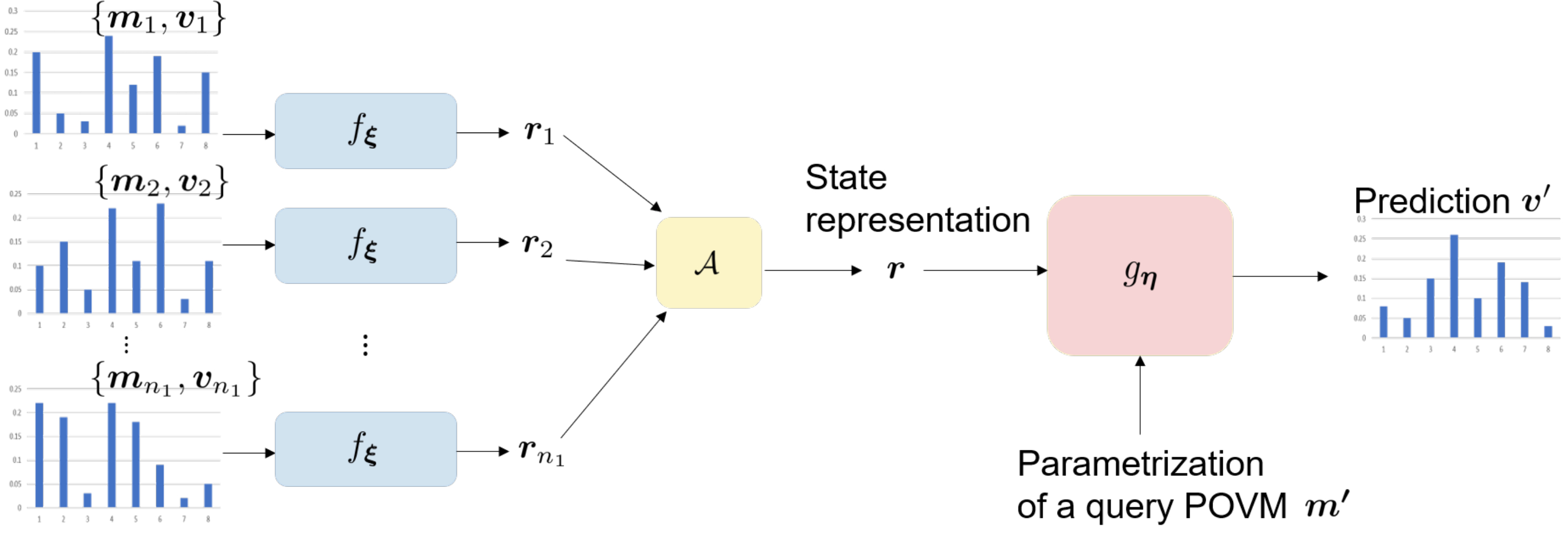}
    \caption{Structure of GQNQ.}
    \label{fig:gqnq_struc}
\end{figure}

The representation network $f_{\bm{\xi}}$ consists of multiple dense layers \cite{aggarwal2018neural}, also called full-connected layers and we depict its structure in Fig.~\ref{fig:rep_struc}. $\bm{\xi}$ contains trainable parameters of all layers.  The input of the representation network is a pair $(\bm{m}_i, \bm{p}_i)$, where $\bm{m}_i$ is parameterization of a POVM measurement and $\bm{p}_i$ is its corresponding measurement outcome probabilities. The output $\bm{r}_i$ can be regarded as an abstract representation of $(\bm{m}_i, \bm{p}_i)$. Here, for simplicity, we just use the average function $\bm{r}:=\frac{1}{s}\sum_{i=1}^s \bm{r}_i$ as the aggregate function but we believe other more sophisticated
architecture such as recurrent neural network \cite{aggarwal2018neural} may achieve better performance, although it will lead to higher requirements for hardware and hyperparameter tuning. 

\begin{figure}[h]
    \centering
    \includegraphics[width=0.65\textwidth]{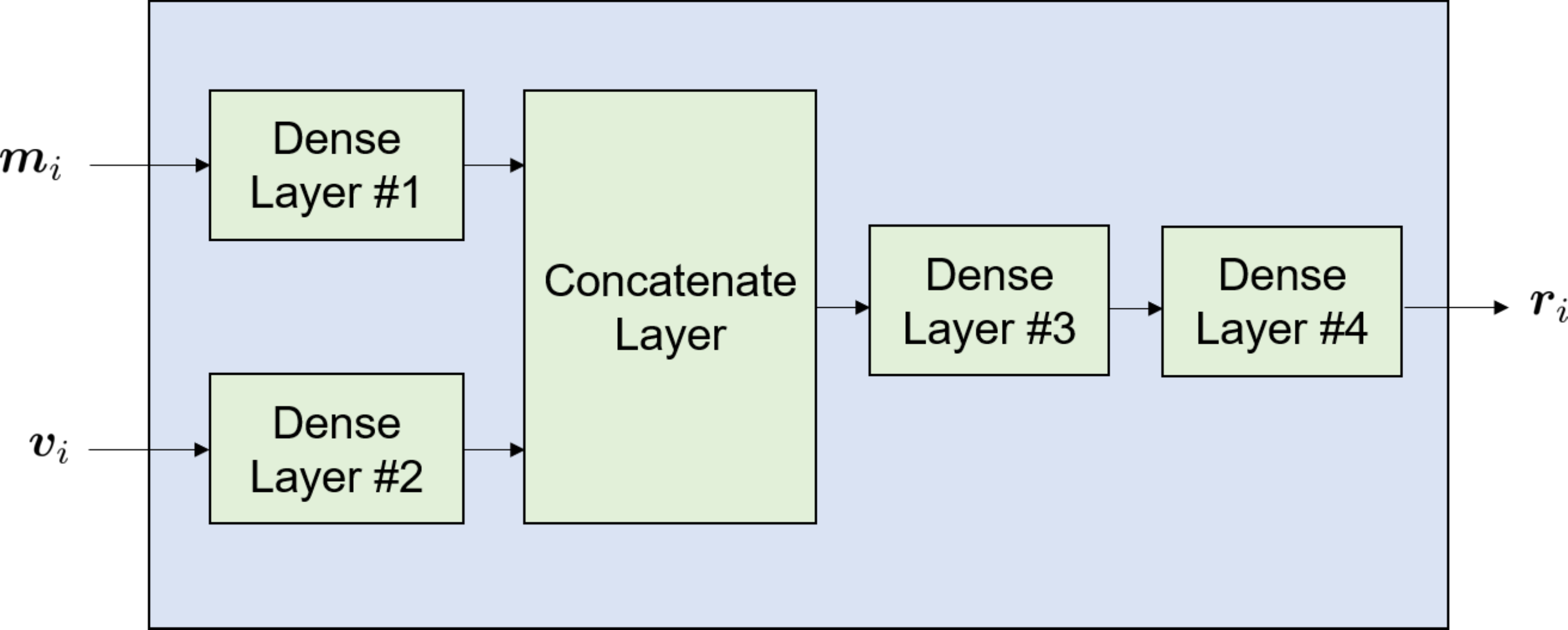}
    \caption{Structure of the representation network.}
    \label{fig:rep_struc}
\end{figure}

The generation network $g_{\bm{\eta}}$ is  special because its structure is different in the training and test phase. Here $\bm{\eta}$ contains all trainable parameters in the generation network. In the test phase, the generation network consists of two dense layers and one long short-term memory (LSTM) cell \cite{hochreiter1997long},  and we depict its structure in Fig.~\ref{fig:gen_struc1}.
The input of this generation network is the state representation $\bm{r}$ and the parameterization $\bm{m}'$ of a query POVM measurement and the output $\mathcal{N}'$ is a distribution of the prediction $\bm{p}'$ of measurement outcome probabilities corresponding to $\bm{m}'$. $\bm{h}_0$, $\bm{c}_0$ and $\bm{u}_0$ are some internal parameters, all of which are initialized as zero tensors. As we can see, the generation network execute Dense Layer \#1 and the LSTM cell for $L$ times while $\bm{m}'$ and $\bm{r}$ are injected to the network for each time.   It is worth mentioning that $\bm{z}_i$ ($i\in\mathbb N ,i< L$) can be viewed as a hidden variable that obeys a prior Gaussian distribution $\mathcal{N}_i$ generated by Dense Layer \#1 from $\bm{h}_i$. In the end, the output $\bm{u}_L$ of the last LSTM cell is fed into a second dense layer (Dense Layer \#2) to obtain the output $\mathcal{N}'$, from which the prediction $\bm{p}'$ is sampled. In Fig.~\ref{fig:gen_struc2}, we depict the generation network exploited in the training. Here, an extra input $\bm{p}'_{true}$ is available because we know the real outcome probabilities in the training phase. Furthermore, we utilize another LSTM cell and another dense layer (Dense Layer \#3) to generate a posterior distribution $\mathcal{N}_i^2$ from $\bm{p}'_{true}$ of the hidden variable $\bm{z}_i$ rather than sampling $\bm{z}_i$ from a prior distribution $\mathcal{N}_i^1$. The advantage of such design is that we can make good use of the information of $\bm{p}'_{true}$ to obtain better $\bm{z}_i$ during the generation. $\bm{h}_0^1$, $\bm{c}_0^1$, $\bm{h}_0^2$, $\bm{c}_0^2$ and $\bm{u}_0$ are some internal parameters, all of which are initialized as zero tensors.

\begin{figure}
    \centering
    \includegraphics[width=0.75\textwidth]{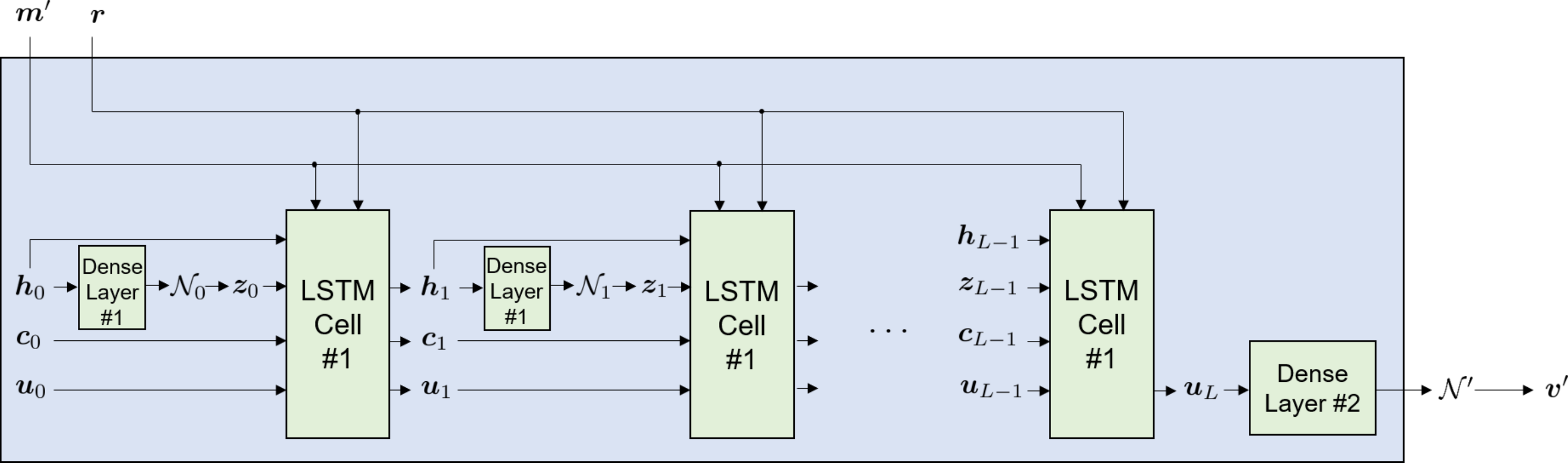}
    \caption{Structure of the generation network in the test.}
    \label{fig:gen_struc1}
\end{figure}

\begin{figure}
    \centering
    \includegraphics[width=0.75\textwidth]{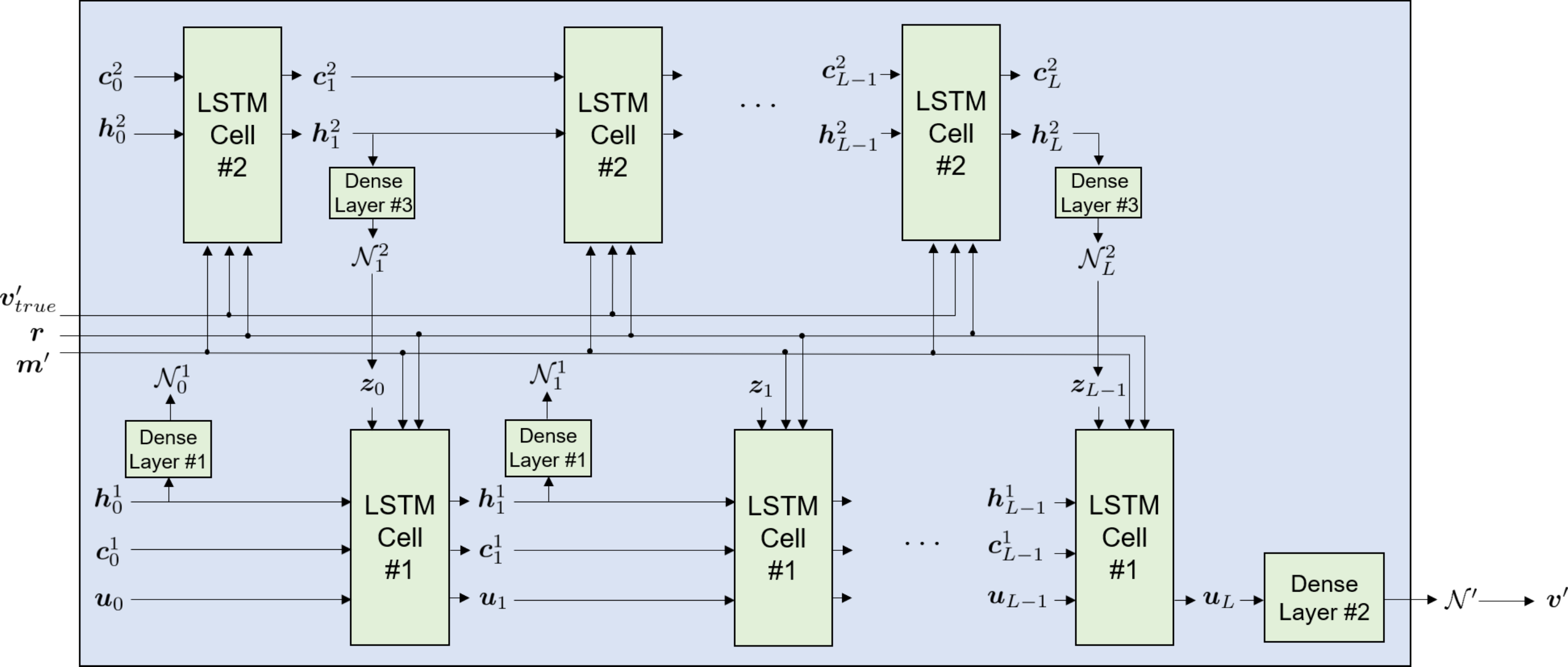}
    \caption{Structure of the generation network in the training.}
    \label{fig:gen_struc2}
\end{figure}

\subsection{Training of GQNQ}
We define the loss function $\mathcal{L}$ of the training in Eq.~(\ref{eqn:loss}). 

\begin{align}\label{eqn:loss}
\mathcal{L}(\bm{\xi},\bm{\eta}) = \mathop{\mathbb{E}}
% \limits_{\{(\bm{m_i},\bm{v_i})\}_{i=1}^{\Vert \mathcal{M}_1\Vert}, \{(\bm{m}',\bm{p}'_{true})\}}
[-\ln\mathcal{N}'(\bm{p}'_{true})+\sum_{j=0}^{L-1}{\rm KL}(\mathcal{N}_j^1,\mathcal{N}_{j+1}^2)],
\end{align}
where $\mathcal{N}'(\bm{p}'_{true})$ denotes the relative likelihood that the variable following distribution $\mathcal{N}'$ takes the value of $\bm{p}'_{true}$ and  ${\rm KL}$ represents KL divergence \cite{kullback1997information} of two Gaussian distributions. The first term of this loss function can be interpreted as the reconstruction loss, which can guide the model to acquire more accurate predictions. The second term is a regularization term utilized for seeking a better prior distribution of the hidden variable $\bm{z}_i$ in the generation process, which is constructive to improving the accuracy of the predictions further.

We adopt batch gradient descent \cite{ruder2016overview} and the Adam optimizer \cite{kingma2014adam} to minimize this loss function in the training. The batch size is set to $10$ or $20$ in all of our experiments and the learning rate decreases gradually with the increase of the number of training epochs.

Our neural networks are implemented by the pytorch \cite{paszke2019pytorch} framework and trained on four NVIDIA GeForce GTX 1080 Ti GPUs. %We exploit the Adam optimizer \cite{kingma2014adam} and batch gradient descent \cite{ruder2016overview} technique to perform the training. 
The training time is less than three hours for each task discussed in this paper.
The ground states of the Hamiltonian of one-dimensional Ising models utilized in our numerical experiments are solved by the exact method for the scenario of $L=6$ and are approximately solved by density-matrix renormalization group (DMRG) \cite{schollwock2005density} for the scenario of $L = 10$, $20$ and $50$. The data of continuous-variable quantum states are generated by simulation tools provided in Strawberry Fields \cite{killoran2019strawberry}. 

We present the whole training procedure by pseudocode in Algorithm \ref{algo:training_GQNQ} following the notations introduced in main text.

\begin{algorithm}
\caption{Training of generative query network for quantum state learning.}\label{algo:training_GQNQ}
% \DontPrintSemicolon
\KwData{number of states in training set $N$, state measurement results $\{\{(\bm{m}_i, \bm{p}^k_i)\}_{i=1}^{n}\}_{k=1}^{N}$, maximum number of known POVM measurement results for each state $a$, maximum number of epochs $E$, learning rate $\delta$, batch size $B$.}

Initialize parameters $\bm{\xi}$ and $\bm{\eta}$ randomly, $e = 0$\;
\While{$e<E$}{
    $\mathcal{L} = 0$\;
    \For{$k=1$ \KwTo $N$}{
        Generate a random integer number $n_1$ from $[1,a]$\;
        Randomly select $n_1$ pairs of $(\bm{m}_i, \bm{p}_i^k)$ from $\{(\bm{m}_i, \bm{p}^k_i)\}_{i=1}^{n}$ and denote them as $\{(\bm{m}_{i_j}, \bm{p}^k_{i_j})\}_{j=1}^{n_1}$, where $\{i_j\}_{j=1}^{n}$ is a permutation of $\{1,\dots,n\}$ \;
        Input each of $\{(\bm{m}_{i_j}, \bm{p}^k_{i_j})\}_{j=1}^{n_1}$ into the representation network $f_{\bm{\xi}}$  to obtain the representations  $\{\bm{r}_{i_j}\}_{j=1}^{n_1}$ as $\bm{r}_{i_j} =f_{\bm{\xi}}(\bm{m}_{i_j}, \bm{p}^k_{i_j})$  \;
        Calculate the state representation by an aggregate function $\mathcal{A}$ as $\bm{r} = \mathcal{A}(\{\bm{r}_{i_j}\}_{j=1}^{n_1})$ \;
        Input $\bm{r}$ and the remaining $\{\bm{m}_{i_j}\}_{j=n_1+1}^{n}$ into the generation network $g_{\bm{\eta}}$ to obtain the predictions $\{\bm{p}'^k_{i_j}\}_{j=n_1+1}^{n}$ of measurement outcome distributions  as $\bm{p}'^k_{i_j} = g_{\bm{\eta}}(\bm{r}, \bm{m}_{i_j})$\;
        Calculate the loss $l$  with Eq.~(\ref{eqn:loss}) by comparing $\{\bm{p}'^k_{i_j}\}_{j=n_1+1}^{n}$ with $\{\bm{p}^k_{i_j}\}_{j=n_1+1}^{n}$ and update $\mathcal{L}$ as $\mathcal{L} = \mathcal{L} + l$ \;
        \If{$k\ \text{mod}\ B = 0$}{
            Calculate $\nabla_{\bm{\xi}} \mathcal{L}$ and $\nabla_{\bm{\eta}} \mathcal{L}$ \;
            Update $\bm{\xi}$ and $\bm{\eta}$ as $\bm{\xi} = \bm{\xi}-\delta \nabla_{\bm{\xi}} \mathcal{L}$, $\bm{\eta} = \bm{\eta}-\delta \nabla_{\bm{\eta}} \mathcal{L}$  \;
            $\mathcal{L} = 0$\;
        }
    }
    $e = e + 1$ \; 
    } 
\end{algorithm}

\subsection{Details of Experiments}
\paragraph{Datasets.} In the experiments for ground states of Ising models and XXZ models, the training set is composed of $40$ different states for each $J$ or $\Delta$ while the test set is composed of $10$ different states for each $J$ or $\Delta$. As for the experiments for GHZ state with local rotation and W state with local rotation, we generate $800$ states for training and $200$ states for test. In the experiments for continuous-variable quantum states, we randomly generate $10000$ different cat states, gaussian states and GKP states and split them into training and test sets with $4:1$ ratio.
\paragraph{Number of trainable parameters.} We mainly adopt three kinds of models for three different tasks. In the experiments for learning discrete quantum states, we exploit the models with $6676544$ trainable parameters for the scenario of $L=6$ while exploiting the models with $45484$ trainable parameters for the scenario of $L=10, 20$ and $50$. In the experiments for learning continuous-variable quantum states, we exploit the models with $35572$ trainable parameters. 
\paragraph{Maximum number of known POVM measurement results for each state in the training.} We set maximum number of known POVM measurement results for each state $a$ in the training as $200$ for the six-qubit cases, $50$ for the $10$-, $20$- and $50$-qubit cases and $150$ for the cases of continuous states.
\paragraph{Initialization and learning rate.} For each task, we initialize the parameters of the models randomly before the training. The learning rate is set as $0.01$ initially and decreases as the number of iterations increases. 
\paragraph{Number of epochs and training time.} We usually set the maximum number of epochs $E$ as $200$ and the batch size $B$ as $30$ in the training. The training time varies with the size of training set for each task while the training time is always less than three hours in all of the experiments.

\section{Hyperparameters}\label{sec:hyperparameter}
As introduced above, there are some hyperparameters in our GQNQ model and the most significant ones are the dimensions of $\bm{r}_i$, $\bm{h}_i$ and $\bm{z}_i$, because they affect the size of the state representation and the complexity of the model, and thus affect the performance of the model. We denote them as $d_{\bm{r}}$, $d_{\bm{h}}$ and $d_{\bm{z}}$ respectively. In this section, we conduct a series of experiments to explore how the choice of hyperparameters affects the performance of the proposed model. We take the settings of learning $6$-qubit states introduced in the main text as examples. In each experiment, different  settings of hyperparameters are adopted and the results are shown in Table~\ref{tab:hyper}. 

We can easily find that as the complexity of the model increases, the performance of the model becomes better. However, it must be pointed out that the complexity of the model cannot be arbitrarily 
high considering the memory size and the difficulty of training, and the models with $d_{\bm{r}} = 32$, $d_{\bm{h}} = 96$ and $d_{\bm{z}} = 32$ are the most complicated ones we consider in this paper.

\begin{table}[]
    \centering
        \caption{Average classical fidelity between predicted outcome statistics and real outcome statistics, averaged over all the test states and random query measurements. 
    The eight rows correspond to eight different scenarios, where GQNQ is trained and tested over measurement data of nine sets of states.
    The values of $d_{\bm{r}}$, $d_{\bm{h}}$ and $d_{\bm{z}}$ are different for each column.
    }
    \resizebox{\textwidth}{!}{
    \begin{tabular}{c|c|c|c|c|c|c}
    \hline\hline
       Types of states  & $d_{\bm{r}} = 2$, $d_{\bm{h}} = 2$, $d_{\bm{z}} = 2$
  & $d_{\bm{r}} = 2$, $d_{\bm{h}} = 6$, $d_{\bm{z}} = 2$ & $d_{\bm{r}} = 4$, $d_{\bm{h}} = 12$, $d_{\bm{z}} = 4$ & $d_{\bm{r}} = 8$, $d_{\bm{h}} = 24$, $d_{\bm{z}} = 8$ & $d_{\bm{r}} = 16$, $d_{\bm{h}} = 48$, $d_{\bm{z}} = 16$ & $d_{\bm{r}} = 32$, $d_{\bm{h}} = 96$, $d_{\bm{z}} = 32$ 
  
  \\
        \hline
        (i) Ising ground states with ferromagnetic bias & 0.7987 & 0.8835 & 0.8981 & 0.9255 & 0.9543 & 0.9870\\
        (ii) Ising  ground states with antiferromagnetic bias & 0.7896 & 0.8739 & 0.8894 & 0.9236 & 0.9562 & 0.9869\\
        (iii) Ising ground states with no bias  & 0.7999 & 0.8911 & 0.8993 & 0.9277 & 0.9596 & 0.9895 \\
        (iv) XXZ ground states with ferromagnetic bias   & 0.6386 & 0.7683 & 0.9038 & 0.9546 & 0.9603 & 0.9809\\
        (v) XXZ ground states with XY phase bias  & 0.7515 & 0.8102 & 0.8359 & 0.8924 & 0.9352 & 0.9601\\
        (vi) (i)-(v)  together  & 0.7143 & 0.7709 & 0.8376 & 0.8739 & 0.9178 & 0.9567\\
        (vii) GHZ state with local rotations  & 0.8342 & 0.8816 & 0.9271 & 0.9502 & 0.9579 & 0.9744 \\
        (viii) W state with local rotations  & 0.9249 & 0.9310 & 0.9579 & 0.9733 & 0.9771 & 0.9828 \\
        (ix) (i)-(v), (vii) and (viii) together  & 0.6936 & 0.7685 & 0.8369 & 0.8725 & 0.9085 & 0.9561 \\
        % (x) Arbitrary 6-qubit state & 0.8879 & 0.8879  & 0.8879 & 0.8879 & 0.8879 & 0.8879 & 0.8879 \\

        \hline\hline
    \end{tabular}}
    \label{tab:hyper}
\end{table}

\section{Arbitrary State Learning}\label{sec:randomState}
Furthermore, we conducted experiments to learn arbitrary $6$-qubit quantum states and the results are shown in Table~\ref{tab:arbit}. We claim that all models failed in this case, since we find that they always yield distributions close to the uniform distribution for any query measurement, which means that the model cannot learn an effective state representation to generate accurate measurement outcome statistics. A possible explanation is that the model is not complicated enough to handle an unstructured, highly complex dataset. Although a more complicated model might be more effective intuitively, such a model may require larger training set and be less efficient. As expected, our GQNQ model is designed for quantum states sharing a common structure and is not suitable for arbitrary quantum states.

\begin{table}[h]
    \centering
        \caption{Average classical fidelity between predicted outcome statistics and real outcome statistics for the arbitrary states, averaged over all the test states and random query measurements. 
    The values of $d_{\bm{r}}$, $d_{\bm{h}}$ and $d_{\bm{z}}$ are different for each column.
    }
    \resizebox{\textwidth}{!}{
    \begin{tabular}{c|c|c|c|c|c|c|c}
    \hline\hline
       Types of states & Uniform distribution & $d_{\bm{r}} = 2$, $d_{\bm{h}} = 2$, $d_{\bm{z}} = 2$
  & $d_{\bm{r}} = 2$, $d_{\bm{h}} = 6$, $d_{\bm{z}} = 2$ & $d_{\bm{r}} = 4$, $d_{\bm{h}} = 12$, $d_{\bm{z}} = 4$ & $d_{\bm{r}} = 8$, $d_{\bm{h}} = 24$, $d_{\bm{z}} = 8$ & $d_{\bm{r}} = 16$, $d_{\bm{h}} = 48$, $d_{\bm{z}} = 16$ & $d_{\bm{r}} = 32$, $d_{\bm{h}} = 96$, $d_{\bm{z}} = 32$ 
  
  \\
        \hline
        
        Arbitrary 6-qubit state & 0.8879 & 0.8879  & 0.8879 & 0.8879 & 0.8879 & 0.8879 & 0.8879 \\

        \hline\hline
    \end{tabular}}
    \label{tab:arbit}
\end{table}

\section{Generalization from Informationally Incomplete Measurements}\label{sec:generalization}

In this section, we will further discuss the generalization performance of our proposed model in the examples of six-qubit quantum states. We mainly focus on how the information completeness of the measurement class affects the final performance. 
%In this section, we mainly focus on how the information completeness of the measurement class affects the final performance. 
Rather than setting the class of measurements $\mathcal{M}$ as the set of all $729$ six-qubit Pauli-basis measurements, we construct $\mathcal{M}$ by randomly selecting $72$ different six-qubit Pauli-basis measurements in each experiment here. For each dataset we discussed, we did  such experiments and averaged the results. The results are shown in Table~\ref{tab:general}. 

As the experimental results show, our proposed model still has a satisfactory performance when the measurement class is not informationally complete. Meanwhile, we also find that the model will generalize worse as the complexity of datasets increases. A possible explanation is that more information is needed to yield accurate state representations when the dataset is composed of multiple types of states.

\begin{table}[h]
    \centering
        \caption{Average classical fidelity between predicted outcome statistics and real outcome statistics, averaged over all the test states and random query measurements. The measurement class $\mathcal{M}$ is composed of $72$ different six-qubit Pauli-basis measurements in the case of informationally incomplete measurements. 
    }
    % \resizebox{\textwidth}{23mm}{
    \begin{tabular}{c|c|c}
    \hline\hline
       Types of states &  Informationally complete $\mathcal{M}$ &  Informationally incomplete $\mathcal{M}$
  
  \\
        \hline
        (i) Ising ground states with ferromagnetic bias &  0.9870 & 0.9865\\
        (ii) Ising  ground states with antiferromagnetic bias & 0.9869 & 0.9863\\
        (iii) Ising ground states with no bias & 0.9895 & 0.9812 \\
        (iv) XXZ ground states with ferromagnetic bias & 0.9809  & 0.9713\\
        (v) XXZ ground states with XY phase bias & 0.9601 &  0.9495\\
        (vi) (i)-(v)  together & 0.9567 & 0.9447\\
        (vii) GHZ state with local rotations & 0.9744 & 0.9694 \\
        (viii) W state with local rotations & 0.9828 &  0.9824 \\
        (ix) (i)-(v), (vii) and (viii) together & 0.9561 & 0.9442\\

        \hline\hline
    \end{tabular}%}
    \label{tab:general}
\end{table}

We also study the generalization performances of GQNQ for continuous-variable states when $\mathcal M$ is information incomplete over the truncated subspace of interest (less than $30$ photons). In the main text, $\mathcal M$ consists of $300$ homodyne measurements with equidistant phases $\theta$ and is informationlly complete over the truncated subspace with less than $300$ photons~\cite{leonhardt1997measuring}. In contrast, here we test the scenario where $\mathcal M$ consists of only $10$ homodyne measurement settings with phases $\theta\in\{0, \pi/10, \dots, 9\pi/10 \}$, and $\mathcal S$ is subset of $\mathcal M$ containing $5$ random homodyne measurement settings. Note that now $\mathcal M$ is insufficient to fully characterize a density matrix on a truncated subspace with more than nine photons. We train and test GQNQ using data from three types of continuous-variable states as discussed in the main text in this scenario.
The average and worst classical fidelities over all query measurements, together with the comparison with the scenario where $|\mathcal M|=300$ and $|\mathcal S|=10$, are presented in Table~\ref{tab:cvgeneral}.
The results show that even in this information incomplete scenario, GQNQ still shows great prediction performance on all the types of test states. Again this is because the states we consider fall within lower-dimensional corners of the subspace with limited photons.

\begin{table}[]
    \centering
        \caption{  Generalization performances of GQNQ on continuous-variable quantum states.  
    }
    \begin{tabular}{c | c |c|c|c}
    \hline\hline
       Type of states for training and test  & $|\mathcal M|=300$ (Avg) & $|\mathcal M|=300$ (Worst) &  $|\mathcal M|=10$ (Avg) & $|\mathcal M|=10$ (Worst) \\
        \hline
        (i) Squeezed thermal states &  $0.9973 $ & $0.9890$ & $0.9953$ & $0.9901$ \\
        (ii) Cat states  & $0.9827 $ & $0.9512$ & $0.9571$ & $0.8920$ \\
        % (1) and (2) together & $0.9878 $ & $0.9604$  & $0.9753$ & $0.9391$ & $0.9874$ & $0.9575$  \\
        (iii) GKP states & $0.9762 $ & $0.9405$  & $0.9633$ & $0.9470$ \\
        (iv) (i)-(iii) together & $0.9658$ & $0.9077$  & $0.9507$ & $0.8843$ \\

        \hline\hline
    \end{tabular}
    \label{tab:cvgeneral}
\end{table}

\section{Overfitting}\label{sec:overfitting}

Overfitting due to unbalanced data can be an important issue when GQNQ is used for learning across multiple types of states. Here we study the six-qubit scenario where GQNQ is trained and tested on the union of the datasets of ground states of Ising model, ground states of XXZ model, GHZ states and W states with local rotations. Specifically, one type of states is chosen to be underrepresented, appearing $10$ times less frequently than any other type of states in the whole training dataset. Then we test the prediction performances of GQNQ with respect to both this underrepresented type of states and the other types of states as shown in Fig.~\ref{fig:overfitting}.

The results show that the performance of GQNQ with unbalanced training data depends on the state under consideration. For W states with local rotations we find that unbalanced training data has little effect on the performance. The situation is similar for the ground states of the Ising model in the ferromagnetic phase. In contrast, the prediction for XXZ model in the XY phase drops to $0.73$ when the  training data are unbalanced.  The results agree with the phenomenon that the ground states of XXZ model in the XY phase are more difficult to learn than any other type of states we considered.

\begin{figure}
    \centering
    \includegraphics[width=\textwidth]{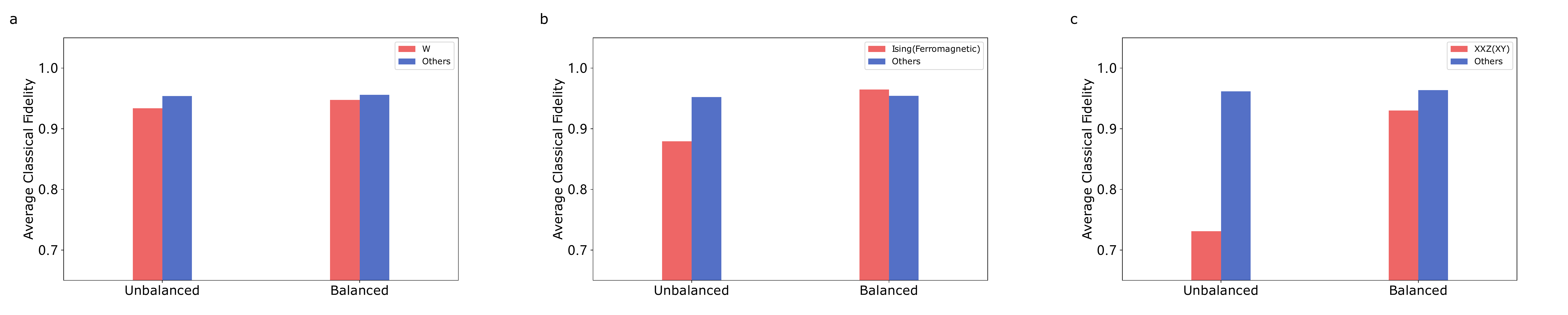}

    \caption{ Performances of GQNQ when training data from different types of quantum states are unbalanced.  %{\color{blue} [say what are the blue and red bars in the histogram.] } 
    In each figure, the red bar represents the classical fidelity with respect to the chosen underrepresented type of states, and the blue bar represents the classical fidelity averaged over all other types of states. 
    Fig.(a) compares the average classical fidelity for W states with local rotations and the average classical fidelity for all other states when the ratio of the size of training data from W states to any other type is $1:10$ and $1:1$, respectively. Fig.(b) compares the average classical fidelity for ground states of ferromagnetic Ising model with the average classical fidelity for all other states when the ratio of the size of training data from ferromagnetic Ising model to any other type is $1:10$ and $1:1$, respectively. 
    Fig.(c) compares the average classical fidelity for ground states of XXZ model in XY phase with the average classical fidelity for all other states when the ratio of the size of training data from XXZ model in XY phase to any other type is $1:10$ and $1:1$, respectively.}
    \label{fig:overfitting}
\end{figure}

\section{Additional Experiments}\label{sec:additional}
%In addition to the numerical experiments described in the main body of the paper, we also perform other experiments and we will present more results in this section.
\subsection{Ising model}

We study the performance of GQNQ for $10$-, $20$- and $50$-qubit Ising ground states when the measurements are nearest-neighbour two-qubit Pauli measurements. Different from the setting in the main text, here we choose $J_i$ as a Gaussian variable with mean value $J$ and variance $0.01$. Hence when $J$ is around $0$, both ferromagnetic interactions and antiferromagnetic interactions are present with high probability. We find that GQNQ cannot give good predictions of outcome statistics in this scenario when both ferromagnetic and antiferromagnetic interactions exist. The results, together with the comparison with the scenario where each $J_i$ is chosen to be the absolute value (or the opposite of the absolute value, for $J<0$) of the Gaussian variable, are presented in Fig.~\ref{fig:comparisonIsing}.

\begin{figure}
    \centering
    \includegraphics[width=0.95\textwidth]{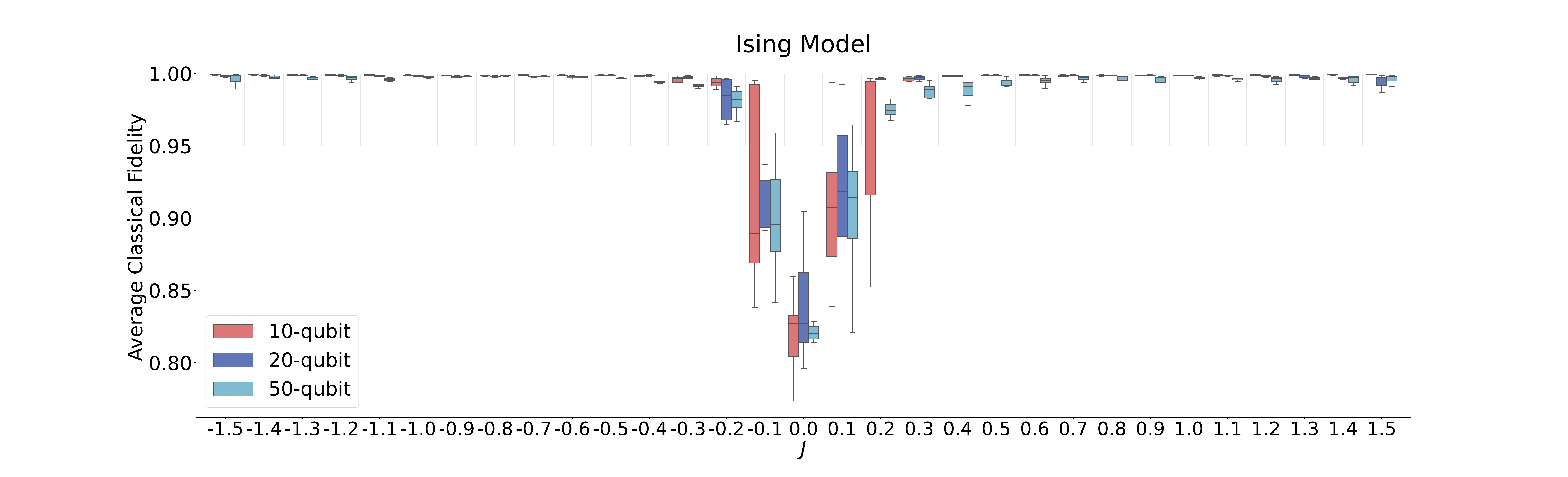}
    \includegraphics[width=0.8\textwidth]{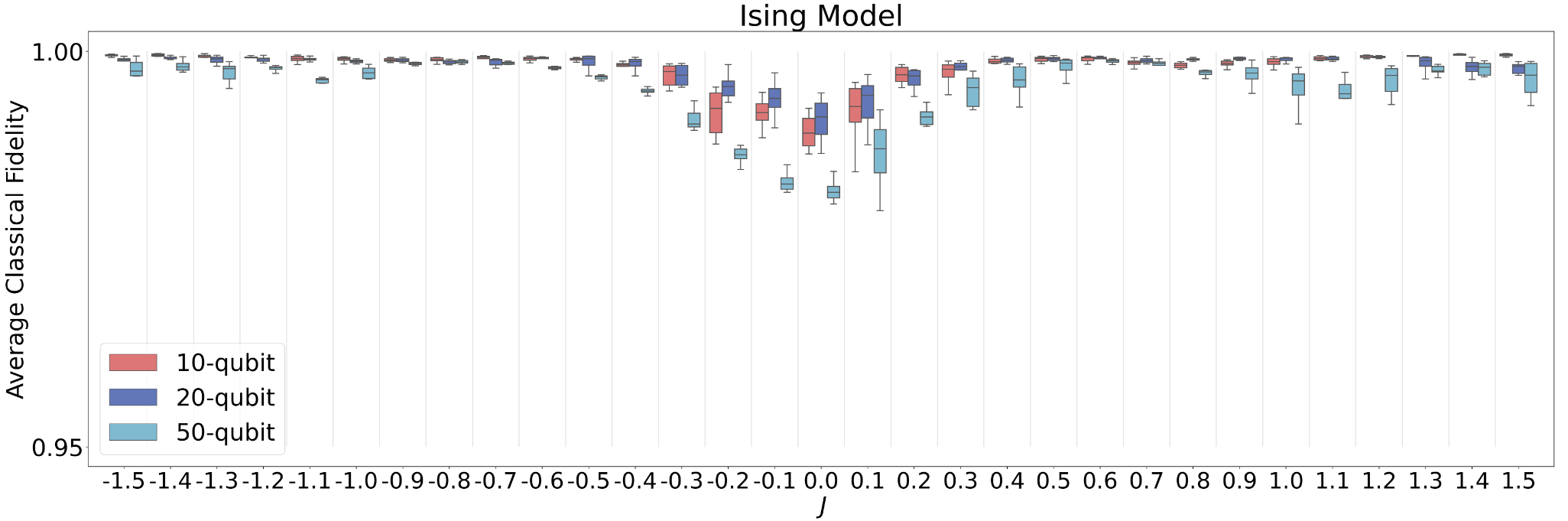}
    \caption{Comparison between the performances of GQNQ in Ising model when both ferromagnetic and antiferromagnetic interactions are present near $J=0$ (top) vs when only either ferromagnetic or antiferromagnetic interactions are present near $J=0$ (bottom).}
    \label{fig:comparisonIsing}
\end{figure}

\subsection{Cat states}
For the numerical experiments on learning of continuous-variable quantum states, we provide an example of
comparison between predictions and ground truths for a cat state in Fig.~\ref{fig:cat} here.

\begin{figure}[h]
    \centering
    \includegraphics[width=0.7\textwidth]{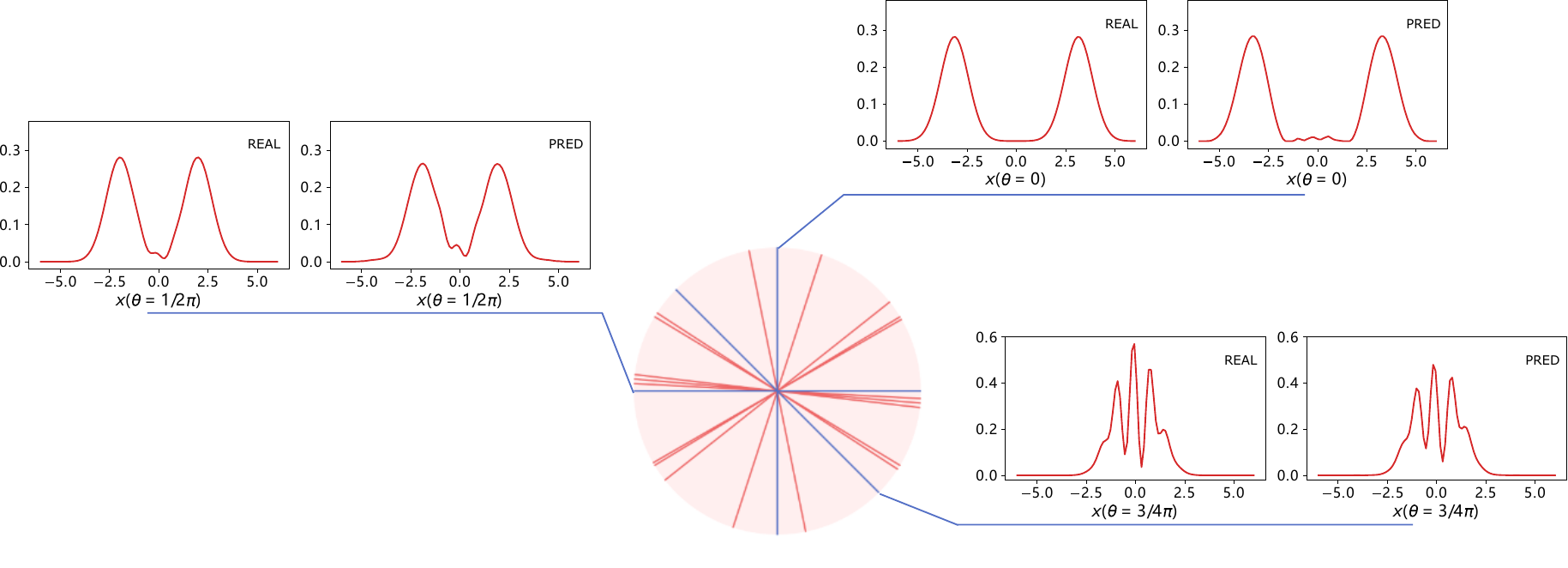}
    \caption{
    The true outcome probability density (left) and the predicted probability density (right) for cat state 
 $\ket{2.22+1.41\text{i}, \pi/4}_{\text{cat}}$ at quadrature phases $\theta=0$, $\theta=\pi/2$ and $\theta=3\pi/4$, respectively, given the measurement outcome densities at ten random quadrature phases.
In the middle circle, ten red lines passing through the center represent those quadrature phases at which measurement outcome statistics are known, and three blue lines passing through the center represent those quadrature phases at which measurement outcome statistics are to be predicted.}
    \label{fig:cat}
\end{figure}

\iffalse
\begin{figure}[h]
    \centering
    \includegraphics[width=0.7\textwidth]{figures/gkp.pdf}
    \caption{
    The real outcome probability density (left) and the predicted probability density (right) for gkp state 
 $\ket{0.15,\pi/24,7\pi/8}_{\text{gkp}}$ at quadrature phases $\theta=0$, $\theta=\pi/2$ and $\theta=3\pi/4$, respectively, given the measurement outcome densities at ten random quadrature phases.
In the middle circle, fifteen red lines passing through the center represent those quadrature phases at which measurement outcome statistics are known, and three blue lines passing through the center represent those quadrature phases at which measurement outcome statistics are to be predicted.}
    \label{fig:gkp}
\end{figure}
\fi

\section{Training with data from the state to be characterized}\label{sec:singleState}

In this section, we discuss how to train our GQNQ model with data only
from the quantum state to be characterized. In this setting, GQNQ behaves as a completely unsupervised learner that predicts the outcome statistics of unperformed
measurements using measurement data obtained from the quantum state under consideration. The set $\mathcal{M}_*$ of 
fiducial measurements in the training coincides with the set $\mathcal S$ of performed measurements. In the training, GQNQ is trained with $s (s < n)$  measurement results  $\{(\bm{m}_i,\bm{p}_i)\}_{i=1}^{s}$ corresponding to $\mathcal S$. When the training is finished, the trained model can be utilized to predict the outcome statistics corresponding to $\mathcal M \setminus \mathcal S$.

We present the whole training procedure in such setting by pseudocode in Algorithm \ref{algo:training_GQNQ_single}. 

\begin{algorithm}[H]
\caption{Training of GQNQ with data provided from the quantum state to be characterized.}\label{algo:training_GQNQ_single}
% \DontPrintSemicolon
\KwData{State measurement results $\{(\bm{m}_i, \bm{p}_i)\}_{i=1}^{s}$ of the quantum state to be characterized corresponding to the set of reference measurements $\mathcal{M}_*$ , maximum number of known POVM measurement results $a(a<s)$ in the training, maximum number of epochs $E$, learning rate $\delta$.}

Initialize parameters $\bm{\xi}$ and $\bm{\eta}$ randomly, $e = 0$\;
\While{$e<E$}{
    $\mathcal{L} = 0$\;
    Generate a random integer number $n_1$ from $[1,a]$\;
    Randomly select $n_1$ pairs of $(\bm{m}_i, \bm{p}_i)$ from $\{(\bm{m}_i,\bm{p}_i)\}_{i=1}^{s}$ and denote them as $\{(\bm{m}_{i_j}, \bm{p}_{i_j})\}_{j=1}^{n_1}$, where $\{i_j\}_{j=1}^{s}$ is a permutation of $\{1,\dots,s\}$ \;
    Input each of $\{(\bm{m}_{i_j}, \bm{p}_{i_j})\}_{j=1}^{n_1}$ into the representation network $f_{\bm{\xi}}$  to obtain the representations  $\{\bm{r}_{i_j}\}_{j=1}^{n_1}$ as $\bm{r}_{i_j} =f_{\bm{\xi}}(\bm{m}_{i_j}, \bm{p}_{i_j})$  \;
    Calculate the state representation by an aggregate function $\mathcal{A}$ as $\bm{r} = \mathcal{A}(\{\bm{r}_{i_j}\}_{j=1}^{n_1})$ \;
    Input $\bm{r}$ and the remaining $\{\bm{m}_{i_j}\}_{j=n_1+1}^{s}$ into the generation network $g_{\bm{\eta}}$ to obtain the predictions $\{\bm{p}'_{i_j}\}_{j=n_1+1}^{s}$ of measurement outcome distributions  as $\bm{p}'_{i_j} = g_{\bm{\eta}}(\bm{r}, \bm{m}_{i_j})$\;
    Calculate the loss $l$  with Eq.~(\ref{eqn:loss}) by comparing $\{\bm{p}'_{i_j}\}_{j=n_1+1}^{s}$ with $\{\bm{p}_{i_j}\}_{j=n_1+1}^{s}$ and update $\mathcal{L}$ as $\mathcal{L} = \mathcal{L} + l$ \;
    
    Calculate $\nabla_{\bm{\xi}} \mathcal{L}$ and $\nabla_{\bm{\eta}} \mathcal{L}$ \;
    Update $\bm{\xi}$ and $\bm{\eta}$ as $\bm{\xi} = \bm{\xi}-\delta \nabla_{\bm{\xi}} \mathcal{L}$, $\bm{\eta} = \bm{\eta}-\delta \nabla_{\bm{\eta}} \mathcal{L}$  \;
    $\mathcal{L} = 0$\;

    $e = e + 1$ \; 
    } 
\end{algorithm}

\end{widetext}

\end{document}